%% file: TGR.tex
\documentclass[copyright]{eptcs}
\providecommand{\event}{TERMGRAPH 2018 Post-proceedings}
\makeatletter
\@ifundefined{lhs2tex.lhs2tex.sty.read}%
  {\@namedef{lhs2tex.lhs2tex.sty.read}{}%
   \newcommand\SkipToFmtEnd{}%
   \newcommand\EndFmtInput{}%
   \long\def\SkipToFmtEnd#1\EndFmtInput{}%
  }\SkipToFmtEnd

\newcommand\ReadOnlyOnce[1]{\@ifundefined{#1}{\@namedef{#1}{}}\SkipToFmtEnd}
\usepackage{amstext}
\usepackage{amssymb}
\usepackage{stmaryrd}
\DeclareFontFamily{OT1}{cmtex}{}
\DeclareFontShape{OT1}{cmtex}{m}{n}
  {<5><6><7><8>cmtex8
   <9>cmtex9
   <10><10.95><12><14.4><17.28><20.74><24.88>cmtex10}{}
\DeclareFontShape{OT1}{cmtex}{m}{it}
  {<-> ssub * cmtt/m/it}{}

\DeclareFontShape{OT1}{cmtt}{bx}{n}
  {<5><6><7><8>cmtt8
   <9>cmbtt9
   <10><10.95><12><14.4><17.28><20.74><24.88>cmbtt10}{}
\DeclareFontShape{OT1}{cmtex}{bx}{n}
  {<-> ssub * cmtt/bx/n}{}
\newcommand{\Conid}[1]{\mathit{#1}}
\newcommand{\Varid}[1]{\mathit{#1}}
\newcommand{\anonymous}{\kern0.06em \vbox{\hrule\@width.5em}}

\usepackage{polytable}

\@ifundefined{mathindent}%
  {\newdimen\mathindent\mathindent\leftmargini}%
  {}%

\def\resethooks{%
  \global\let\SaveRestoreHook\empty
  \global\let\ColumnHook\empty}
\newcommand*{\savecolumns}[1][default]%
  {\g@addto@macro\SaveRestoreHook{\savecolumns[#1]}}
\newcommand*{\restorecolumns}[1][default]%
  {\g@addto@macro\SaveRestoreHook{\restorecolumns[#1]}}
\newcommand*{\aligncolumn}[2]%
  {\g@addto@macro\ColumnHook{\column{#1}{#2}}}

\resethooks

\newcommand{\onelinecommentchars}{\quad-{}- }
\newcommand{\commentbeginchars}{\enskip\{-}
\newcommand{\commentendchars}{-\}\enskip}

\newcommand{\visiblecomments}{%
  \let\onelinecomment=\onelinecommentchars
  \let\commentbegin=\commentbeginchars
  \let\commentend=\commentendchars}

\newcommand{\invisiblecomments}{%
  \let\onelinecomment=\empty
  \let\commentbegin=\empty
  \let\commentend=\empty}

\visiblecomments

\newlength{\blanklineskip}
\newcommand{\hsindent}[1]{\quad}% default is fixed indentation
\let\hspre\empty
\let\hspost\empty

\EndFmtInput
\makeatother
\ReadOnlyOnce{polycode.fmt}%
\makeatletter

\newcommand{\hsnewpar}[1]%
  {{\parskip=0pt\parindent=0pt\par\vskip #1\noindent}}

\newcommand{\hscodestyle}{}

\newcommand{\sethscode}[1]%
  {\expandafter\let\expandafter\hscode\csname #1\endcsname
   \expandafter\let\expandafter\endhscode\csname end#1\endcsname}

  {\par\noindent
   \advance\leftskip\mathindent
   \hscodestyle
   \let\\=\@normalcr
   \let\hspre\(\let\hspost\)%
   \pboxed}%
  {\endpboxed\)%
   \par\noindent
   \ignorespacesafterend}

  {\hsnewpar\abovedisplayskip
   \advance\leftskip\mathindent
   \hscodestyle
   \let\hspre\(\let\hspost\)%
   \pboxed}%
  {\endpboxed%
   \hsnewpar\belowdisplayskip
   \ignorespacesafterend}

  {\hsnewpar\abovedisplayskip
   \advance\leftskip\mathindent
   \hscodestyle
   \let\\=\@normalcr
   \(\pboxed}%
  {\endpboxed\)%
   \hsnewpar\belowdisplayskip
   \ignorespacesafterend}

\newcommand{\plainhs}{\sethscode{plainhscode}}

\plainhs

  {\hsnewpar\abovedisplayskip
   \advance\leftskip\mathindent
   \hscodestyle
   \let\\=\@normalcr
   \(\parray}%
  {\endparray\)%
   \hsnewpar\belowdisplayskip
   \ignorespacesafterend}

  {\parray}{\endparray}

  {\(\parray}{\endparray\)}

\def\codeframewidth{\arrayrulewidth}
\RequirePackage{calc}

  {\parskip=\abovedisplayskip\par\noindent
   \hscodestyle
   \arrayrulewidth=\codeframewidth
   \tabular{@{}|p{\linewidth-2\arraycolsep-2\arrayrulewidth-2pt}|@{}}%
   \hline\framedhslinecorrect\\{-1.5ex}%
   \let\endoflinesave=\\
   \let\\=\@normalcr
   \(\pboxed}%
  {\endpboxed\)%
   \framedhslinecorrect\endoflinesave{.5ex}\hline
   \endtabular
   \parskip=\belowdisplayskip\par\noindent
   \ignorespacesafterend}

\newcommand{\framedhslinecorrect}[2]%
  {#1[#2]}

  {\(\def\column##1##2{}%
   \let\>\undefined\let\<\undefined\let\\\undefined
   \newcommand\>[1][]{}\newcommand\<[1][]{}\newcommand\\[1][]{}%
   \def\fromto##1##2##3{##3}%
   }{\) }%

  {\let\orighscode=\hscode
   \let\origendhscode=\endhscode
   \def\endhscode{\def\hscode{\endgroup\def\@currenvir{hscode}\\}\begingroup}
   \orighscode\def\hscode{\endgroup\def\@currenvir{hscode}}}%
  {\origendhscode
   \global\let\hscode=\orighscode
   \global\let\endhscode=\origendhscode}%

\makeatother
\EndFmtInput
\ReadOnlyOnce{agda.fmt}%

\RequirePackage[T1]{fontenc}
\RequirePackage[utf8x]{inputenc}
\RequirePackage{ucs}
\RequirePackage{amsfonts}

\DeclareUnicodeCharacter{737}{\textsuperscript{l}}
\DeclareUnicodeCharacter{8718}{\ensuremath{\blacksquare}}
\DeclareUnicodeCharacter{8759}{::}
\DeclareUnicodeCharacter{9669}{\ensuremath{\triangleleft}}
\DeclareUnicodeCharacter{8799}{\ensuremath{\stackrel{\scriptscriptstyle ?}{=}}}
\DeclareUnicodeCharacter{10214}{\ensuremath{\llbracket}}
\DeclareUnicodeCharacter{10215}{\ensuremath{\rrbracket}}

\renewcommand\Varid[1]{\mathord{\textsf{#1}}}
\let\Conid\Varid
\newcommand\Keyword[1]{\textsf{\textbf{#1}}}
\EndFmtInput

\usepackage{MnSymbol}
\usepackage{stmaryrd}
\usepackage{wasysym} % \pointer
\usepackage{graphicx}

\let\mathscr=\mathcal % trying to avoid too many math alphabets
\DeclareUnicodeCharacter{401}{{\j\kern-0.258em F}} % U+0191
\DeclareUnicodeCharacter{428}{{\kern0.25em\j\kern-0.46em T}} % U+01A2
\def\RELconversesymb{\smallsmile}
\def\RELconv#1{#1^{%
     \mkern-1mu{}{\raise0.3ex\hbox{\tiny$\RELconversesymb$}}}%
    \kern-0.1em{}}
\DeclareUnicodeCharacter{728}{\ensuremath{\RELconv{}}} % U+02D8
\DeclareUnicodeCharacter{738}{\ensuremath{^{\mathrm{s}}}} % U+02E2
\DeclareUnicodeCharacter{739}{\ensuremath{{}^{*}}} % U+02E3:
\DeclareUnicodeCharacter{7506}{\ensuremath{{}^{o}}}  % U+1D52:
\DeclareUnicodeCharacter{7510}{\ensuremath{{}^{p}}}  % U+1D56:
\DeclareUnicodeCharacter{785}{\ensuremath{{}^{\smallfrown}}} % U+0311
\DeclareUnicodeCharacter{915}{\ensuremath{\Gamma}} % \textGamma
\DeclareUnicodeCharacter{926}{\ensuremath{\Xi}}
\DeclareUnicodeCharacter{934}{\ensuremath{\Phi}}
\DeclareUnicodeCharacter{936}{\ensuremath{\Psi}}
\DeclareUnicodeCharacter{937}{\ensuremath{\Omega}}

\DeclareUnicodeCharacter{945}{\ensuremath{\alpha}} % \textalpha
\DeclareUnicodeCharacter{946}{\ensuremath{\beta}} % \textbeta
\DeclareUnicodeCharacter{947}{\ensuremath{\gamma}} % \textgamma
\DeclareUnicodeCharacter{948}{\ensuremath{\delta}} % \textdelta
\DeclareUnicodeCharacter{953}{\ensuremath{\iota}}
\DeclareUnicodeCharacter{954}{\ensuremath{\kappa}}
\DeclareUnicodeCharacter{955}{\ensuremath{\lambda}} % U+03BB \lambda
\DeclareUnicodeCharacter{960}{\ensuremath{\pi}}
\DeclareUnicodeCharacter{961}{\ensuremath{\rho}}
\DeclareUnicodeCharacter{966}{\ensuremath{\phi}} % \textphi
\DeclareUnicodeCharacter{968}{\ensuremath{\psi}} % \textpsi

\DeclareUnicodeCharacter{7506}{\ensuremath{{}^{\mathsf{o}}}} % U+1D52
\DeclareUnicodeCharacter{7510}{\ensuremath{{}^{\mathsf{p}}}} % U+1D56
\DeclareUnicodeCharacter{7515}{\ensuremath{{}^{\mathsf{v}}}}  % U+1D5B ᵛ
\DeclareUnicodeCharacter{7522}{\ensuremath{{}_{i}}}  % U+1D62
\DeclareUnicodeCharacter{7523}{\ensuremath{{}_{r}}}  % U+1D63
\DeclareUnicodeCharacter{8336}{\ensuremath{{}_{a}}} % U+2090
\DeclareUnicodeCharacter{8337}{\ensuremath{{}_{e}}}  % U+2091
\DeclareUnicodeCharacter{8338}{\ensuremath{{}_{o}}}  % U+2092
\DeclareUnicodeCharacter{8341}{\ensuremath{{}_{h}}} % U+2095
\DeclareUnicodeCharacter{8343}{\ensuremath{{}_{l}}} % U+2097
\DeclareUnicodeCharacter{8344}{\ensuremath{{}_{m}}} % U+2098
\DeclareUnicodeCharacter{8345}{\ensuremath{{}_{n}}} % U+2099
\DeclareUnicodeCharacter{8346}{\ensuremath{{}_{p}}} % U+209A
\DeclareUnicodeCharacter{8347}{\ensuremath{{}_{s}}}  % U+209B
\DeclareUnicodeCharacter{11388}{\ensuremath{{}_{j}}} % U+2C7C
\DeclareUnicodeCharacter{8348}{\ensuremath{{}_{t}}}  % U+209C
\DeclareUnicodeCharacter{8227}{\ensuremath{\filledmedtriangleright}} % U+2023 \unfinished
\DeclareUnicodeCharacter{8252}{\ensuremath{{!\kern-0.2em!}}} % U+203C:
\DeclareUnicodeCharacter{8263}{\ensuremath{{?\kern-0.2em?}}} % U+2047:
\DeclareUnicodeCharacter{8271}{\ensuremath{\fcmp}} % 8271 = U+204F: % REVERSED SEMICOLON

\DeclareUnicodeCharacter{10814}{\ensuremath{\fcmp}} % U+2A3E  \fcmp

\DeclareUnicodeCharacter{8499}{\ensuremath{\mathcal{M}}} % U+2133 curly M
\DeclareUnicodeCharacter{8475}{\ensuremath{\mathcal{R}}} % U+211B curly R
\DeclareUnicodeCharacter{8495}{{\it{e}}} % U+212F curly e
\DeclareUnicodeCharacter{8594}{\ensuremath{\rightarrow}} % U+2192 \to

\DeclareUnicodeCharacter{8599}{\ensuremath{\nearrow}} % U+2197
\DeclareUnicodeCharacter{8608}{\ensuremath{\twoheadrightarrow}} % U+21A0
\DeclareUnicodeCharacter{8611}{\ensuremath{\rightarrowtail}} % U+21A3
\DeclareUnicodeCharacter{8614}{\ensuremath{\mapsto}} %U+21A6:
\DeclareUnicodeCharacter{8618}{\ensuremath{\lhookrightarrow}} %U+21AA:
\DeclareUnicodeCharacter{8634}{\ensuremath{\rcurvearrowne}} % U+21BA
\DeclareUnicodeCharacter{8636}{\ensuremath{\leftharpoonup}} % U+21BC:
\DeclareUnicodeCharacter{8638}{\ensuremath{\upharpoonright}} % U+21BE:
\DeclareUnicodeCharacter{8639}{\ensuremath{\upharpoonleft}} % U+21BF:
\DeclareUnicodeCharacter{8642}{\ensuremath{\downharpoonleft}} % U+21C2:
\DeclareUnicodeCharacter{8643}{\ensuremath{\downharpoonleft}} % U+21C3:
\DeclareUnicodeCharacter{8649}{\ensuremath{\rightrightarrows}} % U+21C9 RIGHTWARDS PAIRED ARROWS

\DeclareUnicodeCharacter{8656}{\ensuremath{\Leftarrow}} % U+21D0:
\DeclareUnicodeCharacter{8663}{\ensuremath{\mathbf{\Nearrow}}} % U+21D7:
\DeclareUnicodeCharacter{8669}{\ensuremath{\rightlsquigarrow}} % U+21DD:
\DeclareUnicodeCharacter{8674}{\ensuremath{\dashedrightarrow}} %U+21E2
\DeclareUnicodeCharacter{8680}{\hbox{\pointer}} % 8680 U+21E8
\DeclareUnicodeCharacter{8702}{\ensuremath{{-}\kern-0.24em\smalltriangleright}} % U+21FE
\DeclareUnicodeCharacter{8760}{\ensuremath{\dotminus}} % U+2238
\DeclareUnicodeCharacter{8762}{\ensuremath{{{:}\kern-0.3em{-}\kern-0.3em{:}}}} % U+223A ∺
\DeclareUnicodeCharacter{8763}{\ensuremath{{\kern0.3em{:}\kern-0.3em{\sim}}}} % U+223B ∻
\DeclareUnicodeCharacter{8767}{\hbox{\large$\sim$}} % U+223F:
\DeclareUnicodeCharacter{8788}{\ensuremath{:=}}  % U+2254:
\DeclareUnicodeCharacter{8845}{\ensuremath{\kern-0.06em\scriptstyle{\cupdot}\kern-0.06em}} % U+228D
\DeclareUnicodeCharacter{8869}{\ensuremath{\bot}} % U+22A5 \bot

\DeclareUnicodeCharacter{8889}{\ensuremath{\fivedots}} % U+22B9 % \unfinished
\DeclareUnicodeCharacter{8988}{\ensuremath{\ulcorner}} % U+231C:
\DeclareUnicodeCharacter{8989}{\ensuremath{\urcorner}} % U+231D:
\DeclareUnicodeCharacter{9001}{\ensuremath{\langle}} % U+2329:
\DeclareUnicodeCharacter{9002}{\ensuremath{\rangle}} % U+232A:
\DeclareUnicodeCharacter{9023}{\ensuremath{\RELlresR}} % U+233F:
\DeclareUnicodeCharacter{9024}{\ensuremath{\RELrresR}} % U+2340:
\DeclareUnicodeCharacter{9070}{\ensuremath{\underline{;}}} % U+236E
\DeclareUnicodeCharacter{9312}{\hbox{%
\raise0.3ex\hbox{$\bigcirc$}\kern-0.7em\raise0.1ex\hbox{\small 1}\kern0.3em}} % U+2460
\DeclareUnicodeCharacter{9374}{\hbox{\small(c)}} % U+249E:
\DeclareUnicodeCharacter{9375}{\hbox{\small(d)}} % U+249F:
\DeclareUnicodeCharacter{9384}{\hbox{\small(m)}} % U+24A8:
\DeclareUnicodeCharacter{9386}{\hbox{\small(o)}} % U+24AA:
\DeclareUnicodeCharacter{9390}{\hbox{\small(s)}} % U+24AE:
\DeclareUnicodeCharacter{9398}{\hbox{%
\raise0.3ex\hbox{$\bigcirc$}\kern-0.8em\raise0.1ex\hbox{\small A}\kern0.3em}}  % U+24B6
\DeclareUnicodeCharacter{9406}{\hbox{%
\raise0.3ex\hbox{$\bigcirc$}\kern-0.6em\raise0.1ex\hbox{\small I}\kern0.4em}}  % U+24BE
\DeclareUnicodeCharacter{9417}{\hbox{%
\raise0.3ex\hbox{$\bigcirc$}\kern-0.8em\raise0.1ex\hbox{\small T}\kern0.3em}} % U+24C9
\DeclareUnicodeCharacter{9432}{\hbox{%
\raise0.3ex\hbox{$\bigcirc$}\kern-0.6em\raise0.1ex\hbox{\small i}\kern0.3em}} % U+24D8
\DeclareUnicodeCharacter{9443}{\hbox{%
\raise0.3ex\hbox{$\bigcirc$}\kern-0.7em\raise0.1ex\hbox{\small t}\kern0.3em}} % U+24D8
\DeclareUnicodeCharacter{9585}{\ensuremath{\RELlres}} % U+2571:
\DeclareUnicodeCharacter{9586}{\ensuremath{\RELrres}} % U+2572:
\DeclareUnicodeCharacter{9645}{\ensuremath{{\scriptstyle{\sqsubset}\kern-0.5em{\sqsupset}}}} % U+25AD:
\DeclareUnicodeCharacter{9646}{\kern-0.1em{\vrule height1.57ex depth0.43ex width0.12em}\kern-0.1em} % U+25AE:
\DeclareUnicodeCharacter{9647}{\ensuremath{{[}\kern-0.2em{]}}} % U+25AF:
\DeclareUnicodeCharacter{9658}{\ensuremath{\filledtriangleright}} % U+25BA:
\DeclareUnicodeCharacter{9668}{\ensuremath{\filledtriangleleft}} % U+25C4:
\DeclareUnicodeCharacter{9671}{\ensuremath{\smalllozenge}} % U+25C7:
\DeclareUnicodeCharacter{9703}{\ensuremath{\filledsquare\kern-0.2em\smallsquare}} % U+25E7
\DeclareUnicodeCharacter{9704}{\ensuremath{\smallsquare\kern-0.2em\filledsquare}} % U+25E8
\DeclareUnicodeCharacter{10003}{\ensuremath{\checkmark}} % U+2713:
\DeclareUnicodeCharacter{10098}{\ensuremath{\lgroup}} %  U+2772 % \unfinished
\DeclareUnicodeCharacter{10099}{\ensuremath{\rgroup}} %  U+2773 % \unfinished
\DeclareUnicodeCharacter{10138}{\ensuremath{\mathbf{\nearrow}}} % U+279A:
\DeclareUnicodeCharacter{10150}{\ensuremath{\rhookrightarrow}} % U+27A6
\DeclareUnicodeCharacter{10153}{\ensuremath{\rightslice}} % U+27A9
\DeclareUnicodeCharacter{10169}{\ensuremath{\nearrowtail}} % U+27B9
\DeclareUnicodeCharacter{10218}{\ensuremath{\llangle}} % U+27EA:
\DeclareUnicodeCharacter{10219}{\ensuremath{\rrangle}} % U+27EB:
\DeclareUnicodeCharacter{10224}{\ensuremath{\mathbf{\diamondplus}}} % U+27F0:
\DeclareUnicodeCharacter{10230}{\ensuremath{\longrightarrow}} % U+27F6:
\DeclareUnicodeCharacter{10231}{\ensuremath{\longleftrightarrow}} % U+27F7:
\DeclareUnicodeCharacter{10232}{\ensuremath{\Longleftarrow}} % U+27F8:
\DeclareUnicodeCharacter{10753}{\ensuremath{\bigoplus}} % U+2A01:
\DeclareUnicodeCharacter{11025}{\rotatebox{90}{\ensuremath{\Rsh}}} % U+2B11
\DeclareUnicodeCharacter{12106}{\hbox{\underline{\it{Tree}}}} % U+2F4A
\DeclareUnicodeCharacter{12298}{\ensuremath{\llangle}} % U+300A:
\DeclareUnicodeCharacter{12299}{\ensuremath{\rrangle}} % U+300B:
\DeclareUnicodeCharacter{12308}{\ensuremath{\displaystyle(\kern-0.23em\mid}} % 〔 U+3014:
\DeclareUnicodeCharacter{12309}{\ensuremath{\displaystyle\mid\kern-0.23em)}} % 〕 U+3015:
\DeclareUnicodeCharacter{8600}{\ensuremath{\searrow}} %U+2198:
\DeclareUnicodeCharacter{8601}{\ensuremath{\swarrow}} %U+2199:
\DeclareUnicodeCharacter{8648}{\ensuremath{\upuparrows}} % U+21C8
\DeclareUnicodeCharacter{8650}{\ensuremath{\downdownarrows}} % U+21CA
\DeclareUnicodeCharacter{8779}{\ensuremath{\triplesim}} % U+224B:
\DeclareUnicodeCharacter{8902}{\ensuremath{^*}}
\DeclareUnicodeCharacter{9587}{\ensuremath{\RELsyqInfixOP}} % U+2573:
\DeclareUnicodeCharacter{9659}{\ensuremath{\rhd}} % U+25BB:
\DeclareUnicodeCharacter{9657}{\ensuremath{\smalltriangleright}} % U+25B9:
\DeclareUnicodeCharacter{9661}{\ensuremath{\medtriangledown}} % U+25BD:
\DeclareUnicodeCharacter{9665}{\ensuremath{\medtriangleleft}} % U+25C1:
\DeclareUnicodeCharacter{9667}{\ensuremath{\smalltriangleleft}} % U+25C3:
\DeclareUnicodeCharacter{9700}{\ensuremath{\blacktriangledown}} % U+25E4:
\DeclareUnicodeCharacter{9701}{} % U+25E45
\DeclareUnicodeCharacter{9733}{\ensuremath{\star}} % U+2605:
\DeclareUnicodeCharacter{65281}{\ensuremath{\mathbf{!}}} % U+FF01
\DeclareUnicodeCharacter{10754}{\ensuremath{\bigotimes}} % U+2A02
\DeclareUnicodeCharacter{10797}{\rotatebox[origin = c]{270}{$\uplus$}}
\DeclareUnicodeCharacter{10798}{\rotatebox[origin = c]{90}{$\uplus$}}
\DeclareUnicodeCharacter{10809}{\hbox{$\largetriangleup\kern-0.8em+\kern0.2em$}}
\DeclareUnicodeCharacter{12297}{\ensuremath{\large\rangle}} % U+3009:
\DeclareUnicodeCharacter{65339}{\ensuremath{\large[}} % U+FF3B:
\DeclareUnicodeCharacter{119914}{\ensuremath{\mathbf{C}}}
\DeclareUnicodeCharacter{119923}{\ensuremath{\mathbf{L}}}
\DeclareUnicodeCharacter{119964}{\ensuremath{\mathcal{A}}}
\DeclareUnicodeCharacter{119966}{\ensuremath{\mathcal{C}}}
\DeclareUnicodeCharacter{119977}{\ensuremath{\mathcal{N}}}
\DeclareUnicodeCharacter{119982}{\ensuremath{\mathcal{S}}}
\DeclareUnicodeCharacter{119995}{\ensuremath{\mathbf{f}}}
\DeclareUnicodeCharacter{119999}{\ensuremath{\mathbf{j}}}
\DeclareUnicodeCharacter{120018}{\ensuremath{\mathcal{C}}}
\DeclareUnicodeCharacter{120019}{\ensuremath{\mathcal{D}}}
\DeclareUnicodeCharacter{120020}{\ensuremath{\mathcal{E}}}
\DeclareUnicodeCharacter{120021}{\ensuremath{\mathcal{F}}}
\DeclareUnicodeCharacter{120022}{\ensuremath{\mathcal{G}}}
\DeclareUnicodeCharacter{120023}{\ensuremath{\mathcal{H}}}
\DeclareUnicodeCharacter{120024}{\ensuremath{\mathcal{I}}}
\DeclareUnicodeCharacter{120025}{\ensuremath{\mathcal{J}}}
\DeclareUnicodeCharacter{120026}{\ensuremath{\mathcal{K}}}
\DeclareUnicodeCharacter{120027}{\ensuremath{\mathcal{L}}}
\DeclareUnicodeCharacter{120028}{\ensuremath{\mathcal{M}}}
\DeclareUnicodeCharacter{120033}{\ensuremath{\mathcal{R}}}
\DeclareUnicodeCharacter{120793}{\ensuremath{\triv}} % U+1D7D9:
\renewcommand{\hscodestyle}{\small}

\makeatletter
\def\mkcommand#1{\expandafter\gdef\csname #1\endcsname}
\makeatother

\def\titlerunning{Semantics-Preserving DPO-Based Term Graph Rewriting}
\title{%
Semantics-Preserving DPO-Based Term Graph Rewriting%
  \thanks{%
    This research is supported by the National Science and
    Engineering Research Council of Canada, NSERC.
  }%thanks
}

\def\authorrunning{Wolfram Kahl \& Yuhang Zhao}
\author{Wolfram Kahl
\institute{%
McMaster University,
Hamilton, Ontario, Canada,
}
\email{kahl@cas.mcmaster.ca}
\and Yuhang Zhao
\institute{%
McMaster University,
Hamilton, Ontario, Canada,
}
\email{zhaoy36@mcmaster.ca}
}

\date{\today}
\usepackage{etex}

       \usepackage{graphicx}

\def\sectionname{Sect.\null{}}
\usepackage{sectref}

\usepackage{amsthm}

\theoremstyle{definition}
\newtheorem{Def}{Definition}[section]

\usepackage{theref}
\def\TheRefWithPageRef#1{}

\newref{Def}{Def.\null{}}

\newdef{Conv}{Convention}
\newdef{The}{Theorem}
\newdef{Lem}{Lemma}
\newdef{Prop}{Proposition}
\newref{Prop}{Prop.\null{}}
\newdef{Cor}{Corollary}
\newdef{Fact}{Fact}
\newdef{Example}{Example}
\newref{Fig}{Fig.\null{}}

\def\fcmp{\,{\scriptstyle\fatsemi}\,}

\def\tfun{\mathop{\rightarrow}}

\long\def\ignore#1{}

\usepackage{CalcStyleV12}

\usepackage{objA}
\usepackage{RelNotation}
\RELsquareSymbols

\renewcommand{\arraystretch}{1.1}

\input diagram-2.2

\usepackage{squareDiags}

\def\etal{\textsl{et al.\null{}}}

\usepackage{edcomms}
\makeatletter
\def\@listI{\leftmargin\leftmargini
            \labelsep 0.3em
            \labelwidth\leftmargini
            \advance\labelwidth-\labelsep
            \parsep 0\p@ \@plus1\p@ \@minus\p@
            \topsep 3\p@ \@plus1\p@ \@minus2\p@
            \itemsep1\p@}
\let\@listi\@listI
\@listi
\makeatother
\def\nat{\mathbb{N}}

\usepackage{formac}
\usepackage{qed}
\let\QED=\qed

\usepackage{CalABC}
\let\phi=\varphi

\def\Cleft{\big\{\;}\def\Cright{\;\big\}}

\def\BMAkern{\kern-5.7pt}

\usepackage{dsfont}
\def\triv{\mathds{1}} %% Works with xelatex!

\def\Obj{\ensuremath{\Conid{Obj}}}
\def\Mor{\ensuremath{\Conid{Mor}}}
\def\src{\ensuremath{\Varid{src}}}
\def\trg{\ensuremath{\Varid{trg}}}

\def\catC{{\cat C}}

\def\catCo{\catC_{0}}

\def\SetCat{\mbox{\emph{Set\/}}}

\def\CThom#1#2#3{\Mor_{#1}[#2,#3]}

\makeindex
\def\ELab{\ensuremath{\Conid{ELab}}}%

\def\categ#1{\mbox{\textbf{#1}}} % Avoiding ``Too many math alphabets used in version normal.''
\def\catC{\categ{C}}
\def\catCo{\catC_{0}}
\Tdef{Nabla}{\nabla}
\TdefA{NablaU}{#1}{\nabla_{\kern-.2ex{#1}}}
\def\exch{\mathbb{X}}
\def\exchU#1{\mathbb{X}_{#1}}
\def\munit{\triv}
\let\origRELcomp\RELcomp
\let\RELcomp\fcmp % for gs-monoidal definitions
\def\CGpicRaw#1#2{\includegraphics#2{drawings/CG/#1}}
\def\CGpic#1{\CGpicRaw{#1}{[scale=0.29]}}
\def\DHG#1#2{\ensuremath{\Conid{DHG}}_{#1,#2}}
\def\TG#1#2{\ensuremath{\Conid{TG}}_{#1,#2}}
\def\arity{\ensuremath{\Varid{arity}}}
\def\length{\ensuremath{\Varid{length}}}
\def\Fin#1{\ensuremath{\Conid{Fin}}_{#1}}
\def\Label{\mathcal{L}}
\def\Edge{\mathcal{E}}
\def\Inner{\mathcal{I}}
\def\Node{\mathcal{N}}
\def\eOut{\ensuremath{\Varid{eOut}}}
\def\gOut{\ensuremath{\Varid{gOut}}}
\def\eIn{\ensuremath{\Varid{eIn}}}
\def\eLabel{\ensuremath{\Varid{eLabel}}}
\def\sem#1{\lsem #1\rsem}
\usepackage{hyperref}
\begin{document}
\maketitle

%{{{ abstract
\begin{abstract}
% TERMGRAPH 2018 extended abstract submission:
Term graph rewriting is important as ``conceptual implementation''
of the execution of functional programs,
and of data-flow optimisations in compilers.
One way to define term graph transformation rule application
is via the well-established and intuitively accessible double-pushout (DPO)
approach; we present a new result proving semantics preservation
for such DPO-based term graph rewriting.
\end{abstract}
%}}}

%{{{ %%%% Keywords:
%%%% \begin{keywords}
%%%% Term graph transformation
%%%% Double-pushout graph transformation
%%%% Semantics preservation
%%%% \end{keywords}
%}%keywords
%}}}

%{{{ \section{Introduction and Related Work}
\section{Introduction and Related Work}

%\citet{Assmann-2000} Graph Rewrite Systems for Program Optimization

%{{{ Term graph rewriting goes back to Wadsworth
Term graph rewriting goes back to % the PhD thesis of
Wadsworth \cite{Wadsworth-1971}, who proposed it as an efficient implementation
mechanism for the $\lambda$-calculus.
This aspect has remained dominant in the term graph literature;
for example, Rose \cite{Rose-1993} %uses term graph rewriting to
defines an operational semantics of a lazy functional programming
language
via term graph rewriting;
%Also on the operational semantics side,
Ariola, Klop and Plump \cite{Ariola-Klop-Plump-2000} study confluence of
term graph rewriting using bisimilarity.
%}}}
%
%{{{ Implementation of Term Rewriting
When justifying term graph rewriting as a correct implementation
technique (for, in particular, functional programming),
most of the literature approaches this from the relationship with term
rewriting.
For example,
when Plump \cite{Plump-2002} writes about ``Essentials of Term Graph Rewriting'',
soundness and completeness are considered only with respect to term
rewriting.
%% 
%% \citet{Toyama-Smetsers-vanEekelen-Plasmeijer-1993}
%% ``The Functional Strategy and Transitive Term Rewriting Systems''
%% %(special class of strongly sequential systems \cite{Huet-Levy-1979})
%% \edcomm{WK}{Anything to do with term graphs?}
%% 
Kennaway \etal{} \cite{Kennaway-Klop-Sleep-deVries-1993c,Kennaway-Klop-Sleep-deVries-1994}
% The Adequacy of Term Graph Rewriting for Simulating Term Rewriting
define a notion of simulation to prove adequacy of
term graph rewriting for finite and rational term rewriting.
%\citet{Kennaway-Klop-Sleep-deVries-1994} study adequacy of rewriting
%of cyclic term graphs for simulating rewriting of infinite terms.
%}}}

\medskip
%{{{ Attempting DPO
When attempting to employ traditional categorial approaches to graph rewriting,
the so-called ``algebraic approach'', to term graph rewriting,
two main problems arise: First, categories of ``standard'' term graph
homomorphisms typically do not have all pushouts,
since unification translates into pushouts,
and second, the interface graphs needed both for the double-pushout (DPO)
approach and for the single-pushout approach (to capture the domain of
morphisms) are typically not term graphs,
but some kind of ``term graphs with holes''.
Term graph rewriting is therefore a niche  of graph transformation
that has pioneered exploration of formalisms
where pushout squares are generalised in some way,
in particular by using different morphisms in the horizontal and
vertical directions of the standard DPO drawing.
%}}}

%{{{ heterogeneous squares
For example,
Banach \cite{Banach-1993}
%{A Fibration Semantics for Extended Term Graph Rewriting}
defines ``DACTL'' term graph rewriting using a modified opfibration, and
Kahl \cite{Kahl-1996,Kahl-1997b} uses both fibrations and opfibrations to define rewriting of
term graphs with variable binding.
A different approach to using separate classes of horizontal and
vertical morphisms for term graph rewriting has been proposed by
Duval \etal{} \cite{Duval-Echahed-Prost-2009},
%``A Heterogeneous Pushout Approach to Term-Graph Transformation''
who are using a specific rule concept as morphisms
in the horizontal direction in their ``heterogeneous pushout approach''.
More recently, motivated by % transformation of
attributed graphs,
which share some characteristics with term graphs,
Habel and Plump \cite{Habel-Plump-2012} propose
``{$\mathcal{M},\mathcal{N}$}-adhesive transformation systems''
as one general framework to accommodate different classes of morphisms
in the horizontal and vertical directions of the double-pushout setting.
%}}}

%\citet{Corradini-Rossi-1993}
%``A New Term Graph Rewriting Formalism: Hyperedge Replacement Jungle Rewriting''
%\unfinished

\medbreak
%{{{ Modelling Term Graphs via GS-Monoidal Categories
%% \subsection*{Modelling Term Graphs via GS-Monoidal Categories}
Corradini and Gadducci \cite{Corradini-Gadducci-1999-APTG,Corradini-Gadducci-2002b}
opened up a new way of investigating term graphs
by defining gs-monoidal categories
as a variant of Lawvere theories \cite{Lawvere-1963}.
Gs-monoidal categories are an intermediate concept between
symmetric monoidal categories and cartesian (monoidal) categories;
the only difference with the latter is that,
the ``duplicator'' transformation $\nabla$
producing diagonal maps $\nabla_{A} : A \tfun A \otimes A$
and the ``terminator'' transformation $!$ with components
$!_{A} : A \tfun \triv$
are both \emph{not} assumed to be natural transformations
(that is, for a morphism $F : A \tfun B$,
 the equations $F \,\fcmp\, \nabla_{B} = \nabla_{A} \,\fcmp\, (F \otimes F)$
 and $F \,\fcmp\,  !_{B} = !_{A}$ do \emph{not} necessarily hold.).
 
Corradini and Gadducci demonstrate in
\cite{Corradini-Gadducci-1999-APTG}
that taking natural numbers as objects
and term graphs with $m$ inputs and $n$ outputs as morphisms from object $m$
to object $n$
produces a free gs-monoidal category,
and thus they automatically obtain a functorial semantics for term graphs
in arbitrary gs-monoidal categories, which include
all Cartesian categories, and so in particular also $\SetCat$.
Continuing this line of work,
Corradini and Gadducci obtain
semantics preservation for
a low-level definition of ``ranked dag rewriting''
% Def. 7: ``rule'': single dag with ``left root'' and ``right root'' (as in HOPS)
and involving ``contexts'' analogous to the contexts of term
rewriting
\cite{Corradini-Gadducci-1997,Corradini-Gadducci-1999-cyclic}.
%% \edcomm{WK}{Lemma 22 produces contexts for matchings;
%%    before Lemma 22 the refer to ``full version''
%%    --- is that \citep{Corradini-Gadducci-1999-cyclic}?
%% }%edcomm
% 
% \citet{Corradini-Gadducci-1999-cyclic}
% \unfinished
% 
Finally, in \cite{Corradini-Gadducci-2005} they show a quasi-adhesive category of term
graphs, but emphasise that adhesive categorial rewriting in that
category does not quite match term graph rewriting.
They mention in their conclusion that a possible alternative is
to perform the DPO on a super-category of hypergraphs;
this is essentially the approach we are elaborating here.
%}}}
%
As an example consider \Figref{Fig_example-DPO-rewriting}, showing the application of a rule
corresponding to the term rule
$\ (x_1 + x_2) - x_2 \;\longrightarrow\; x_1\ $
to rewrite a term graph corresponding to
$\ y_1 + ((y_2 + y_3) - y_3) \times y_4\ $
to $\ y_1 + y_2 \times y_4$.

%{{{ {Yuhang/Graphs/example-DPO-rewriting}\Figlabel{Fig_example-DPO-rewriting}
\begin{figure}[h!]
%    \centering
%    {\includegraphics[scale=0.71,viewport=67 79 524 726,clip]{Yuhang/Graphs/example-DPO-rewriting}}
%
\ignore{
%{{{ DIAG using Yuhang/Graphs/example-DPO-rewriting_v1
\vrule height10ex width0pt depth0pt
\DIAGV{210}   
{\fbox{\includegraphics[scale=0.71,viewport=80 507 168 726,clip]{Yuhang/Graphs/example-DPO-rewriting_v1}}\ L\kern-0.7em}
\n {\War{\Phi}} \n
{\fbox{\includegraphics[scale=0.71,viewport=260 507 337 726,clip]{Yuhang/Graphs/example-DPO-rewriting_v1}}\ G\kern-1em}
\n {\Ear{\Psi}} \n
{\fbox{\includegraphics[scale=0.71,viewport=446 507 509 726,clip]{Yuhang/Graphs/example-DPO-rewriting_v1}}\ R\kern-1em}
\diag
\vrule height13ex width0pt depth0pt
\DIAGV{70}   
{\Sar{M_1}} \n {} \n {} \n {} \n {} \n {} \n {\Sar{\Chi}} \n {} \n {} \n {} \n {} \n {} \n {\Sar{M_2}}
\diag
\vrule height23ex width0pt depth0pt
\DIAGV{210}
{\fbox{\includegraphics[scale=0.71,viewport=63 77 194 426,clip]{Yuhang/Graphs/example-DPO-rewriting_v1}}\ A\kern-0.8em}
\n {\War{\Xi}} \n
{\fbox{\includegraphics[scale=0.71,viewport=233 77 359
      426,clip]{Yuhang/Graphs/example-DPO-rewriting_v1}}\ H\kern-1em}
\n {\Ear{\Omega}} \n
{%\strut\kern-0.8em B\
 \fbox{\includegraphics[scale=0.71,viewport=396 77 523 426,clip]{Yuhang/Graphs/example-DPO-rewriting_v1}}\ B\kern-0.5em}
\diag
\vrule height22ex width0pt depth0pt
%}}}
}%ignore
\vrule height9.9ex width0pt depth0pt
%{{{ DIAG using Yuhang/Graphs/example-DPO-rewriting
\DIAGV{190}   
{\fbox{\includegraphics[scale=0.48,viewport=87 565 173 806,clip]{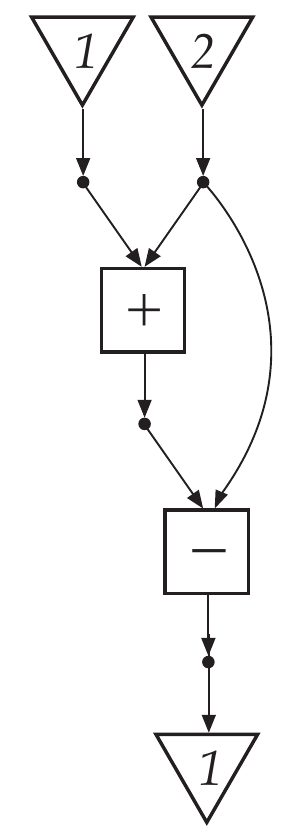}}\ L\kern-0.7em}
\n {\B\War{\Phi}} \n
{\fbox{\includegraphics[scale=0.48,viewport=292 565 372 806,clip]{Yuhang/Graphs/example-DPO-rewriting}}\ G\kern-1em}
\n {\B\Ear{\Psi}} \n
{\fbox{\includegraphics[scale=0.48,viewport=486 565 574 806,clip]{Yuhang/Graphs/example-DPO-rewriting}}\ R\kern-1em}
\diag
\vrule height7.5ex width0pt depth0pt
\DIAGV{63}   
{\Sar{M_1}} \n {} \n {} \n {} \n {} \n {} \n {\Sar{\Chi}} \n {} \n {} \n {} \n {} \n {} \n {\Sar{M_2}}
\diag
\vrule height14ex width0pt depth0pt
\DIAGV{190}
{\fbox{\includegraphics[scale=0.48,viewport=54 97 201 467,clip]{Yuhang/Graphs/example-DPO-rewriting}}\ A\kern-0.8em}
\n {\B\War{\Xi}} \n
{\fbox{\includegraphics[scale=0.48,viewport=250 97 395 467,clip]{Yuhang/Graphs/example-DPO-rewriting}}\ H\kern-1em}
\n {\B\Ear{\Omega}} \n
{%\strut\kern-0.8em B\
 \fbox{\includegraphics[scale=0.48,viewport=440 97 583 467,clip]{Yuhang/Graphs/example-DPO-rewriting}}\ B\kern-0.5em}
\diag
%}}}
\vrule height15.5ex width0pt depth0pt
    \caption{Example term graph rewriting step}\Figlabel{Fig_example-DPO-rewriting}
\end{figure}
%}}}

%}}}

%{{{ \subsection*{Overview}
%\subsection*{Overview}
\medbreak
In \sectref{gsMonCat} we provide details about term graphs and how we
draw them, and the definition of gs-monoidal categories
with explanations how term graphs populate that concept.
In \sectref{TGR-DPO} we present the adaptations we use to obtain a
DPO-based definition of term graph transformation,
and in \sectref{SemPreserve} we sketch the proof that such
transformation steps are semantics-preserving
if the rule sides are semantically equivalent.
%}}}

%{{{ \section{Background: Term Graphs and GS-Monoidal Categories}\sectlabel{gsMonCat}
%\section{Background on GS-Monoidal Categories of Term Graphs}\sectlabel{gsMonCat}
\section{Background: Term Graphs and GS-Monoidal Categories}\sectlabel{gsMonCat}

We are using the ``jungle'' view of term graphs, which goes back to
Hoffmann and Plump \cite{Hoffmann-Plump-1991}
and Corradini and Rossi \cite{Corradini-Rossi-1991},
since this is the view used by the gs-monoidal semantics,
where nodes translate into objects and (hyper-)edges into morphisms.

We assume a set $\Label$ of \emph{edge labels} together with an
  function
  $\arity : \Label \tfun \NN$
  prescribing for each label the number of inputs the corresponding
  edges take. We write $\Fin{k} \defeq \{ i : \NN \with i < k\}$
  for the set containing the first $k$ natural numbers,
  and will use this in particular for the set of graph input nodes.
  
  \begin{Def}\Deflabel{DHG}\Deflabel{TG}
  The set of \emph{directed hypergraph graphs with $m$ inputs and $n$ outputs}
  will be denoted by $\DHG{m}{n}$.
  An element of $\DHG{m}{n}$ is
  a tuple $(\Inner, \Edge, \eLabel, \eOut, \eIn, \gOut)$
  consisting of two sets,
  \begin{itemize}
  \item a set $\Inner$ of \emph{inner nodes},
    from which we construct the set $\Node = \Fin{m} \uplus \Inner$
    of \emph{nodes} as disjoint union of the set
    $\Fin{m}$ of \emph{graph input nodes} and the set $\Inner$ of inner nodes,

  \item a set $\Edge$ of \emph{(hyper-)edges},
  \end{itemize}
  and four functions,
  \begin{itemize}
  \item $\eLabel : \Edge \tfun \Label$ assigning each edge a label,
    \item $\eOut : \Edge \tfun \Inner$ assigning each edge a single
      \emph{edge output node}, which has to be an inner node,
    \item $\eIn : \Edge \tfun \Node^*$ assigning each
      edge a sequence of \emph{edge input nodes},
      which needs to have as its length the arity of the edge's label,
      that is, $\forall e : \Edge \spot \arity (\eLabel(e)) = \length(\eIn(e))$, and
    \item $\gOut : \Fin{n} \tfun \Node$ assigning each output position
      a node.
  \end{itemize}
  A \emph{term graph} is an acyclic directed hypergraph where $\eOut$ is
  bijective;
  we write $\TG{m}{n}$ for the set of term graphs with $m$ inputs and $n$ outputs.
      \qed
\end{Def}

When drawing such hypergraphs and term graphs, we start with the inputs on top
and proceed down to the outputs,
drawing nodes as bullets, and (hyper-)edges as labelled boxes
connected to nodes via (implicitly ordered) input-tentacles and
exactly one output-tentacle.
(Although edges with multiple outputs have uses
for example in the code graphs of
\cite{Kahl-Anand-Carette-2005,Anand-Kahl-2009b},
most of the literature, including all the cited work by Corradini and
Gadducci, only considers single-output operations (edges),
so we also do this here.)
Graph input nodes are declared by attaching a triangle pointing to the
input node --- input nodes are necessarily distinct, and cannot be
output nodes of edges.
Graph input nodes are frequently called ``variable nodes'',
and translated into distinct variables for a term reading.
Graph output nodes (in the literature frequently referred to as
``roots'')
are declared by attaching a triangle pointing away
from them --- any node can be used as a graph output any number of
times.

A graph with multiple graph outputs is interpreted as
standing for a tuple of terms:
The left box in the following drawing depicts a term graph (from $\TG{2}{1}$)
corresponding to the term
\linebreak
``$(x_1 + x_2) * x_2$'',
while the term graph (from $\TG{2}{2}$) in the right box corresponds
to the pair of terms
$``((x_1 + x_2) * x_2, (x_1 + x_2) * x_2)$''
(or, if $\ensuremath{\Keyword{let}}$-definitions are available,
$``\ensuremath{\Keyword{let}}\ z = (x_1 + x_2) * x_2\ \ensuremath{\Keyword{in}}\ (z, z)$''):

\medskip
\centerline{%
\fbox{\vrule height22.5ex width0pt depth1ex\kern0.9ex\CGpic{sixX}\kern0.9ex}
\kern8em
\fbox{\vrule height22.5ex width0pt depth1ex\kern0.9ex\CGpic{sixX2}\kern0.9ex}}

\kern1ex
\noindent
Term graphs % (i.e., jungles)
with sequential composition ($\fcmp$) and parallel composition ($\otimes$)
form a gs-monoidal category
according to Corradini and Gadducci \cite{Corradini-Gadducci-1999-APTG}:
The objects are the natural numbers
(interpreted as numbers of nodes in the graph input interface,
respectively graph output interface),
and term graphs with $m$ inputs and $n$ outputs
are morphisms from $m$ to $n$.

%{{{ {Def}\Deflabel{ssmc}\Deflabel{gs-monoidal}
\begin{Def}\Deflabel{ssmc}\Deflabel{gs-monoidal}
%{{{ {Not}{category}
%\begin{Not}
For a \emph{category}
%${\catC} = (\Obj_{\catC}, \Mor_{\catC}, \src, \trg, \RELid, \RELcomp)$,
$(\Obj, \Mor, \src, \trg, \RELid, \RELcomp)$,
we write $f : \objA \tfun \objB$
instead of $\src(f) = \objA \;\land\linebreak \trg(f) = \objB$;
composition of two morphisms
$f : \objA \tfun \objB$ and
  $g : \objB \tfun \objC$ is written ``$f \RELcomp g$'',
and the identity for object $\objA$ is $\RELid_{\objA}$.
%\qed
%\end{Not}
%}}}

%{{{ A \emph{symmetric strict monoidal category} 
A \emph{symmetric strict monoidal category} \cite{MacLane-1971}
$%\catC =
( \catCo, \otimes, \munit, \exch )$
consists of a category $\catCo$,
a strictly associative monoidal bifunctor $\otimes$
with $\munit$ as its strict unit,
and a transformation $\exch$
that associates with every two objects $\objA$ and $\objB$
an arrow $\exch_{\objA,\objB} : \objA \otimes \objB \tfun \objB \otimes \objA$
with
$\exch_{\munit,\munit} = \RELid_{\munit}$
and:

\smallskip
\strut\hfill$%
%\begin{array}[b]{rcl@@{\hskip1em}rcl}
    (F \otimes G) \RELcomp \exch_{\objC,\objD} =
    \exch_{\objA,\objB} \RELcomp (G \otimes F)
\enskip,\quad\hfill
    \exch_{\objA,\objB} \RELcomp \exch_{\objB,\objA} =
    \RELid_{\objA} \otimes \RELid_{\objB}
\enskip,\quad\hfill%\\[.3ex]
    \exch_{\objA\otimes\objB,\objC} =
    (\RELid_{\objA} \otimes \exch_{\objB,\objC}) \RELcomp
    (\exch_{\objA,\objC} \otimes \RELid_{\objB})
%\enskip,&
%    \exch_{\munit,\munit} &=& \RELid_{\munit}
\enskip.
%\end{array}
$\hfill%\qed
%\unskip
%\end{Def}
%}}}

\smallskip
\noindent
%{{{ %{Def}[gs-monoidal categories]\Deflabel{gs-monoidal}
%\begin{Def}\Deflabel{gs-monoidal}
$%\catC =
( \catCo, \otimes, \munit, \exch, \Nabla, !  )$
is a {\em strict gs-monoidal category} iff
%\begin{itemize}
%\item
 $( \catCo, \otimes, \munit, \exch )$
  is a symmetric strict monoidal category, and
\begin{itemize}
\item
  $!$ associates with every object $\objA$ of $\catCo$
  an arrow $!_{\objA} : \objA \tfun \munit$, and
\item $\Nabla$ associates with every object $\objA$ of $\catCo$
  an arrow $\NablaU{\objA} : \objA \tfun \objA \otimes \objA$,
  such that:
\end{itemize}
%such that $\RELid_{\munit} = !_{\munit} = \NablaU{\munit}$,
%and the following axioms hold:
%
\BCM
\def\arraystretch{1.3}
\begin{array}[b]{l}
    \NablaU{\objA} \RELcomp (\RELid_{\objA} \otimes \NablaU{\objA})
  \sepA{=}
    \NablaU{\objA} \RELcomp (\NablaU{\objA} \otimes \RELid_{\objA})
\qquad
\qquad
    \NablaU{\objA} \RELcomp \exchU{\objA,\objA}
  \sepA{=}
    \NablaU{\objA}
\qquad
\qquad
    \NablaU{\objA} \RELcomp (\RELid_{\objA} \otimes !_{\objA})
  \sepA{=}
    \RELid_{\objA}
\\
    \NablaU{\objA \otimes \objB} \RELcomp
    (\RELid_{\objA} \otimes \exchU{\objB,\objA} \otimes \RELid_{\objB})
  \sepA{=}
    \NablaU{\objA} \otimes \NablaU{\objB}
\qquad
\qquad
    !_{\objA \otimes \objB}
  \sepA{=}
  !_{\objA} \otimes !_{\objB}
\qquad
\qquad
  \RELid_{\munit} = !_{\munit} = \NablaU{\munit}
\ECMAQ
%}}}
\end{Def}
%}}}

\ignore{% %% now in Intro
The definition of gs-monoidal categories places them between
symmetric monoidal categories and cartesian (monoidal) categories;
the only difference with the latter is that,
the ``duplicator'' transformation $\nabla$
producing diagonal maps $\nabla_{A} : A \tfun A \otimes A$
and the ``terminator'' transformation $!$ with components
$!_{A} : A \tfun \triv$
are both \emph{not} assumed to be natural transformations
(that is, for a morphism $F : A \tfun B$,
 the equations $F \,\fcmp\, \nabla_{B} = \nabla_{A} \,\fcmp\, (F \otimes F)$
 and $F \,\fcmp\,  !_{B} = !_{A}$ do \emph{not} necessarily hold.).
}%ignore

\smallskip
\noindent
For term graphs, the lack of naturality of the ``terminator''
transformation $!$
means that \emph{garbage}
(nodes from which no output is reachable) makes a difference,
such as between the two graphs to the left below,
and the lack of naturality of the ``duplicator'' transformation $\nabla$ means that
\emph{sharing} (use of nodes in more than one consumer r\^ole,
that is, as inputs for edges or as graph outputs)
makes a difference,
such as between the two graphs to the right below.
(The words ``garbage'' and ``sharing'' motivate the name ``gs-monoidal''.)

\kern2ex
%{{{ naturalityViolated graphs
\noindent
\strut\hfill
\fbox{\includegraphics[scale=1,viewport=75 40 99 168,clip]{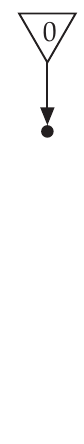}}% !
\hfill
\fbox{\includegraphics[scale=1,viewport=136 40 159 168,clip]{Yuhang/Graphs/naturalityViolated}}% F;!
\hfill
\hfill
\hfill
\fbox{\includegraphics[scale=1,viewport=17 40 39 168,clip]{Yuhang/Graphs/naturalityViolated}}% F
\hfill
\hfill
\hfill
\fbox{\includegraphics[scale=1,viewport=194 40 247 168,clip]{Yuhang/Graphs/naturalityViolated}}% F;\nabla
\hfill
\fbox{\includegraphics[scale=1,viewport=287 40 338 168,clip]{Yuhang/Graphs/naturalityViolated}}% \nabla;FF
\hfill\strut
%}}}
\\[1.2ex]
%{{{ naturalityViolated expressions
\strut\hfill
$\kern0.7em !_A\kern0.7em$
\hfill
$F \ \fcmp\ !_B$
\hfill
\hfill
\hfill
$\kern0.7em F\kern0.7em$
\hfill
\hfill
\hfill
$\kern1.2em F \ \fcmp\ \nabla_B\kern1.2em $
\hfill
$\nabla_A \ \fcmp\  (F \otimes F)\kern-0.5em$
\hfill\strut
%}}}

\kern2ex
\noindent
Corradini and Gadducci \cite{Corradini-Gadducci-1999-APTG} show furthermore
that the term graphs %(jungles)
over a given signature
are arrows of the gs-monoidal category 
freely generated by that signature;
therefore, there always
exists a unique functor
from the gs-monoidal category of term graphs
to any gs-monoidal category.
This induces a functorial semantics for term graphs %(jungles)
in any gs-monoidal category.
(This will frequently be some (cartesian) category of sets,
with some set $\CalV$ chosen as set of \emph{values} ``at a node'';
a term graph with $m$ inputs and $n$ outputs
then has a function of type $\CalV^m \rightarrow \CalV^n$
as semantics.
For code generation applications,
one may construct non-cartesian gs-monoidal semantics categories
where morphisms contain information about resource usage,
such as number of instructions.)
%}}}

%{{{ \section{Adapted DPO for Term Graph Rewriting}\sectlabel{TGR-DPO}
\section{Adapted DPO for Term Graph Rewriting}\sectlabel{TGR-DPO}

We will use the naming of graphs and morphisms used in \Figref{DPO}
for double-square diagrams in the shape of double pushouts.

\begin{figure}[h!]
\kern-0.5ex
\DIAGV{65}
{L} \n {\War{\Phi}} \n {G} \n {\Ear{\Psi}} \n {R}
\nn
{\Sar{M_{\mathrm{1}}}} \n {} \n {\Sar{X}} \n {} \n {\saR{M_{\mathrm{2}}}}
\nn
{A} \n {\B\War{\Xi}} \n {H} \n {\B\Ear{\Omega}} \n {B}
\diag

\kern-1ex
\caption{Naming of objects and morphism in ``DPO-shape'' diagrams}\Figlabel{DPO}
\end{figure}

\smallskip
\noindent
The example term graph transformation step
in our adapted DPO approach
shown in \Figref{Fig_example-DPO-rewriting} in the introduction
in effect closely corresponds to the more low-level
definitions of term graph transformation dominant in the literature:
the ``host graph'' (or ``context graph'') $H$
can be thought of as obtained from
the ``application graph'' $A$
by deleting all edges and inner nodes of $A$
which have a pre-image in $L$,
but no pre-image (via $\Phi \fcmp M_1$) in $G$,
and the ``result graph'' $B$ is obtained from $H$
by ``gluing in'' the right-hand side $R$.

%\noindent
The gluing graph $G$ and the host graph $H$
are obviously not jungles, since they have nodes
that are neither graph input nodes nor edge output nodes,
but they still are directed hypergraphs (DHGs)
in the sense of \Defref{DHG}.
%In fact, we define jungles as \emph{acyclic} DHGs
%where each non-input node is the output node of
%\emph{exactly one} edge. \unfinished

Both for DHGs and for term graphs
we distinguish \emph{matchings},
which preserve edge labelling and incidence structure,
from \emph{homomorphisms}, which in addition preserve
also graph input and output structure:

\begin{Def}
  A \emph{DHG matching} $\Phi = (\Phi_{\Node}, \Phi_{\Edge})$
  from $G_1 : \DHG{m_1}{n_1}$ to $G_2 : \DHG{m_2}{n_2}$
  consists of two functions $\Phi_{\Node} : \Node_1 \tfun \Node_2$
  and $\Phi_{\Edge} : \Edge_1 \tfun \Edge_2$
  satisfying:

\smallskip
  \strut\hfill
  $\eOut_2 \circ \Phi_{\Edge} = \Phi_{\Node} \circ \eOut_1\enskip,$
  \hfill
  $\eLabel_2 \circ \Phi_{\Edge} = \eLabel_1\enskip,$
  \hfill and
  \hfill
  $\eIn_2 \circ \Phi_{\Edge} = \ensuremath{\Varid{map}}\ \Phi_{\Node} \circ \eIn_1$.

\smallskip
\noindent
  A \emph{DHG homomorphism} $\Phi = (\Phi_{\Inner}, \Phi_{\Edge})$
  from $G_1 : \DHG{m}{n}$ to $G_2 : \DHG{m}{n}$
  consists of two functions $\Phi_{\Inner} : \Inner_1 \tfun \Inner_2$
  and $\Phi_{\Edge} : \Edge_1 \tfun \Edge_2$
  such that defining
  $\Phi_{\Node} \defeq Id_{\Fin{m}} \uplus \Phi_{\Inner}$
    turns $(\Phi_{\Node}, \Phi_{\Edge})$
    into a matching from $G_1$ to $G_2$
  and additionally satisfies
  \ $\gOut_2 = \Phi_{\Node} \circ \gOut_1$.

If $G_1$ and $G_2$ are term graphs, then a matching (respectively
homomorphism)
$\Phi$ from $G_1$ to $G_2$ is called a \emph{term graph matching}
(respectively \emph{term graph homomorphism}).
\qed
\end{Def}

\smallskip
The diagram in \Figref{Fig_example-DPO-rewriting} is then a double pushout
in the category of DHG matchings, satisfying the following additional
requirements:

\begin{Def}A DPO diagram in the category of DHG matchings of the shape
  of \Figref{DPO}
  is called a \emph{TG-DPO} iff:
\begin{itemize}
\item $M_1$ and $M_2$ are term graph matchings
  (which implies that $L$, $R$, $A$, and $B$ all are term graphs),
\item $\Phi$, $\Psi$, $\Xi$, $\Omega$ are DHG homomorphisms.
  \qed
\end{itemize}
\end{Def}

\smallskip
\noindent
Superficially, this arrangement looks similar to that of the
$\CalM,\CalN$-adhesive categories of Habel and Plump
\cite{Habel-Plump-2012} ---
we would use DHG homomorphisms for $\CalM$
and term graph matchings for $\CalN$.
However, several of the conditions of $\CalM,\CalN$-adhesive
categories fail to hold for this setting.

The existence of a pushout complement in the category of DHG matchings
is subject to the gluing condition as usual ---
both dangling and identification conflicts can occur.

If the rule $L \Bkar{\Phi} G \Ar{\Psi} R$ consists of
DHG homomorphisms, both the pushout complement construction for the
left square and the pushout construction for the right square will
yield DHG matchings $\Xi$ and $\Omega$ that also respect the graph
interface, and therefore are DHG homomorphisms.

\begin{figure}[h!]
\begin{minipage}[b]{0.42\columnwidth}
    \centering
    %{\includegraphics[scale=0.85]{Yuhang/Graphs/R-DPO-Conflict}}
    %{\includegraphics[scale=1.0,viewport=206 176 403 519,clip]{Yuhang/Graphs/R-DPO-conflict_p78}}
%{{{ DIAG using Yuhang/Graphs/R-DPO-Conflict
\vrule height7ex width0pt depth0pt
\DIAGV{150}   
{\fbox{\includegraphics[scale=0.85,viewport=22 218 44 344,clip]{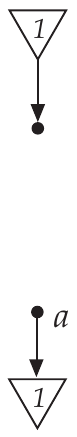}}\ G\kern-1em}
\n {\B\Ear{\Psi}} \n
{\fbox{\includegraphics[scale=0.85,viewport=156 218 181 344,clip]{Yuhang/Graphs/R-DPO-Conflict}}\ R\kern-1em}
\diag
\vrule height7ex width0pt depth0pt
\DIAGV{50}   
{\Sar{\Chi}} \n {} \n {} \n {} \n {} \n {} \n {\Sar{M_2}}
\diag
\vrule height7ex width0pt depth0pt
\DIAGV{150}
{\fbox{\includegraphics[scale=0.85,viewport=22 29 44 154,clip]{Yuhang/Graphs/R-DPO-Conflict}}\ H\kern-1em}
\n {\B\Ear{\Omega}} \n
{%\strut\kern-0.8em B\
 \fbox{\includegraphics[scale=0.85,viewport=137 29 202 154,clip]{Yuhang/Graphs/R-DPO-Conflict}}\ B\kern-0.5em}
\diag
\vrule height8ex width0pt depth0pt
%}}}
    \caption{RHS edge conflict}\Figlabel{Fig_R-DPO-Conflict}
\end{minipage}
\hfill
\begin{minipage}[b]{0.48\columnwidth}
    \centering
    %{\includegraphics[scale=1.1]{Yuhang/Graphs/notMonicMatchingGH}}
    %{\includegraphics[scale=0.85,viewport=208 386 400 721,clip]{Yuhang/Graphs/notMonicMatchingGH_p80}}
%{{{ DIAG using Yuhang/Graphs/notMonicMatchingGH
\vrule height7ex width0pt depth0pt
\DIAGV{150}   
{\fbox{\includegraphics[scale=0.96,viewport=25 227 80 339,clip]{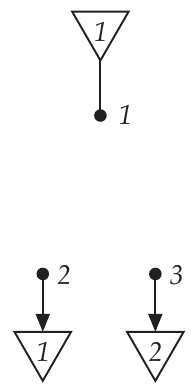}}\ G\kern-1em}
\n {\B\Ear{\Psi}} \n
{\fbox{\includegraphics[scale=0.96,viewport=133 227 188 339,clip]{Yuhang/Graphs/notMonicMatchingGH}}\ R\kern-1em}
\diag
\vrule height7ex width0pt depth0pt
\DIAGV{50}   
{\Sar{\Chi}} \n {} \n {} \n {} \n {} \n {} \n {\Sar{M_2}}
\diag
\vrule height7ex width0pt depth0pt
\DIAGV{150}
{\fbox{\includegraphics[scale=0.96,viewport=25 33 80 146,clip]{Yuhang/Graphs/notMonicMatchingGH}}\ H\kern-1em}
\n {\B\Ear{\Omega}} \n
{%\strut\kern-0.8em B\
 \fbox{\includegraphics[scale=0.96,viewport=133 33 188 146,clip]{Yuhang/Graphs/notMonicMatchingGH}}\ B\kern-1em}
\diag
\vrule height8ex width0pt depth0pt
%}}}
    \caption{Non-injective host matching}\Figlabel{Fig_notMonicMatchingGH}
\end{minipage}
\end{figure}

\noindent
For the right square of the DPO diagram,
we finally have to ensure that $B$ is a term graph,
which is not trivial.
First, the situation shown in
\Figref{Fig_R-DPO-Conflict} would lead to $B$ not being a term graph ---
however, since the $\Phi$-image of node $a$ in $L$
has to be either an input node or the output of an edge,
such a situation cannot occur
at least when the rule LHS $\Phi$ is injective.
(If the image of $a$ is an input node,
then, with $\Phi$ preserving the graph interface,
it cannot be injective.
If the image of $a$ is the output node of an edge in $L$,
then the image in $A$ of that edge needs to be also the image
of the $S$-edge in $H$, which contradicts the left-hand pushout.)
Second, also the example DHG matching pushout in
\Figref{Fig_notMonicMatchingGH}
fails to produce a term graph $B$ --- this situation can be avoided
by restricting the matching $M_1$ to be injective.
(In effect, both constraints together correspond to the restriction to
the ``regular monos'' of \cite[Prop.~4.3]{Corradini-Gadducci-2005}.)

Since the right-hand side $\Psi$ of the rule is a DHG homomorphism,
it is automatically injective on input nodes;
non-injectivity of $\Psi$ therefore can only force identifications that are also
``permissible'' for the host graph,
so we do not need to restrict $\Psi$ to be injective,
which would be highly unwelcome for term graph rewriting.

Therefore, DPOs in the DHG matching category can be used to
rewrite term graphs with rules with injective left-hand sides,
using only injective matchings (which takes care of the identification
part of the gluing condition):
%Altogether, we obtain the following rule applicability theorem:

\begin{The}\Thelabel{TG-DPO}
  Given a term graph rewriting rule $L \Bkar{\Phi} G \Ar{\Psi} R$
  where $L$ and $R$ are term graphs and $\Phi$ and $\Psi$ are
  DHG homomorphisms, with $\Phi$ injective,
  and given further an injective term graph matching $L \Ar{M_1} A$,
  then this setting can be completed to a TG-DPO
  if the dangling condition holds for $M_1$.
  \qed
\end{The}

\smallskip
The fact that $\Phi$ is injective
implies that the output nodes of $L$ are disjoint from the input
nodes;
we call such a term graph \emph{solid}.
%}}}

%{{{ \section{Semantics Preservation of DPO-Transformation of Term Graphs}{SemPreserve}
\section{Semantics Preservation of DPO-Transformation of Term Graphs}\sectlabel{SemPreserve}

While the fact that term graphs form a free gs-monoidal category
gives us semantics of term graphs,
it does not give us semantics of DHGs such as the gluing and host
graphs in most typical rewriting steps.
Rather than trying to artificially obtain some semantics for DHGs
``with holes'', we will transfer the necessary information ``across
the host graph $H$'' at the DHG level.

A starting point could be the decomposition of term graphs
into gs-monoidal expressions as described in
\cite{Corradini-Gadducci-1999-APTG}.
However, instead of extending this expression type
into a type of contexts by including ``placeholders''
as proposed in \cite{Corradini-Gadducci-2002b},
we define contexts at the level of graphs:

\begin{Def}\Deflabel{context}
An \emph{$m,n$-context $(k, A_1, A_2)$ for an $i,j$-parameter} consists of:
\begin{itemize}
\item an \emph{internal interface} object $k$,
\item a \emph{top part} term graph $A_1 : \TG{m}{i + k}$, and
\item a \emph{bottom part} term graph $A_2 : \TG{j + k}{n}$.
\qed
\end{itemize}
\end{Def}

\medskip
\noindent
In the following,
we continue to use ``$\fcmp$'' as sequential composition operator for term graphs,
and ``$\otimes$'' for parallel composition.
Furthermore, ``$\RELid_{k}$'' denotes the \emph{identity}
term graph with $k$ inputs that are also its outputs, in the same
sequence.
The empty DHG with $i$ inputs, and with $j$ distinct output nodes that
are disjoint from the input nodes is written ``$\bot_{i,j}$'';
for the sub-category of DHG \emph{homomorphisms}
restricted to DHGs with $i$ inputs and $j$ outputs,
$\bot_{i,j}$ is the initial object.

\begin{Def}\Deflabel{ImageContextFor}
  An $m,n$-context $(k, A_1, A_2)$ for an $i,j$-parameter is called an
  \emph{image context for} an injective term graph matching $M_1 : L \tfun A$
  starting from term graph $L : \TG{i}{j}$
  iff $\ A \;\cong\; A_1 \fcmp (L \, \otimes\, \RELid_{k}) \,\fcmp \, A_2\ $
  and the nodes and edges of $L$ in that expression precisely constitute
the image of $M_1$ in $A$.
\qed
\end{Def}

By ensuring that there is no ``side entrance'' from within the
application graph $A$
into the image of the LHS $L$, the dangling condition is crucial for the
following result:

\begin{Lem}
Assume a solid term graph $\ L : \TG{i}{j}\ $ to be given,
and let $\ \Phi : \bot_{i,j} \tfun L\ $ be
the (necessarily-injective) DHG homomorphism from $\bot_{i,j}$ to $L$.
If $\ A : \TG{m}{n}\ $ is a term graph
and $\ M_1 : L \tfun A\ $ is an injective term graph matching
that together with $\Phi$ satisfies the dangling condition,
then there is an image context $(k, A_1, A_2)$ for $M_1$.
\qed
\end{Lem}

Such a context can be calculated in several different ways
from the reachability in $A$,
for example by collecting all edges into $A_1$ that are reachable from
the input nodes of $A$ via paths that do not touch the image
of $L$ under $M_1$. The difference $\ (A - A_1) - L\ $ would then induce $A_2$.

\medskip
\def\Ginput{\scalebox{0.8}[1]{\textsf{input}}}%
\def\Goutput{\scalebox{0.8}[1]{\textsf{output}}}%
\def\Gcolimit{\scalebox{0.8}[1]{\textsf{COLIMIT}}}%
Sequential and parallel composition in the gs-monoidal category
of term graphs (as morphisms)
can be obtained as colimits in the category of DHG matchings.
In the following diagram we denote the coproduct injections as $\iota$ and $\kappa$;
for a $X : \DHG{m}{n}$ we use $\Ginput : \RELid_{m} \tfun X$
as the DHG matching mapping $\RELid_{m}$ identically to the input
nodes of $X$, and analogously $\Goutput : \RELid_{n} \tfun X$.

The lower-left box below contains the diagram that has
as its colimit the application graph $A$,
factored into the context $(k, A_1, A_2)$ and an image of the
left-hand side $L$ as
$A\; \cong\; A_1\ \fcmp\ (L\ \otimes\ \RELid_{k})\ \fcmp\ A_2$.

\medskip
%\pagebreak
\noindent
\begin{minipage}{\columnwidth}
%{{{ Factored DPO
\DIAGV{67}
{L} \n {} \n {} \n {\B\Warv{\Phi}{200}} \n {} \n {} \n
{\bot_{i,j}} \n {} \n {} \n {\B\Earv{\Psi}{200}} \n {} \n {} \n {R}
\nn
{\Nbiar{\Gcolimit}{}}
\n {} \n {} \n {} \n {} \n {} \n
{\Nbiar{\Gcolimit}{}}  
\n {} \n {} \n {} \n {} \n {} \n
{\Nbiar{\Gcolimit}{}} 
\diag
\kern0.3ex
\strut\hfill
%{{{ L : i -> j
\fbox{\begin{minipage}[c]{0.18\columnwidth}
\DIAGV{60}
      {\RELid_{i}}
      \nn
      {\Sar{\Ginput}}
      \nn
      {L}
      \nn
      {\Nar{\Goutput}}
      \nn
      {\RELid_{j}}
\diag
\end{minipage}}     
%}}}     
\hfill
$\Bkar{\displaystyle\Phi'}$
\hfill
%{{{ G_0 : i -> j
\fbox{\begin{minipage}[c]{0.18\columnwidth}
\DIAGV{60}
      {\RELid_{i}}
      \nn
      {}
      \nn
      {}
      \nn
      {}
      \nn
      {\RELid_{j}}
\diag
\end{minipage}}     
%}}}     
\hfill
$\Ar{\displaystyle\Psi'}$
\hfill
%{{{ R : i -> j
\fbox{\begin{minipage}[c]{0.18\columnwidth}
\DIAGV{60}
      {\RELid_{i}}
      \nn
      {\Sar{\Ginput}}
      \nn
      {R}
      \nn
      {\Nar{\Goutput}}
      \nn
      {\RELid_{j}}
\diag
\end{minipage}}     
%}}}     
\hfill
\strut
\\[-1.4ex]
\DIAGV{67}
{\Sar{M_1'}}
\n {} \n {} \n {} \n {} \n {} \n
{\Sar{\Chi'}}  
\n {} \n {} \n {} \n {} \n {} \n
{\Sar{M_2'}} 
\diag
\kern1ex
\strut\hfill
%{{{ L -> (k, A_1, A_2)
\fbox{\begin{minipage}[c]{0.23\columnwidth}
\DIAGV{60}
      {} \n {} \n {A_1}
      \nn
      {} \n {} \n {\Nar{\Goutput}}
      \nn
      {\RELid_{i}} \n {\Ear{\iota}} \n {\RELid_{i + k}}
      \nn
      {\Sar{\Ginput}} \n {} \n {\Nar{\kappa}}
      \nn
      {L} \n {} \n {\RELid_{k}}   
      \nn
      {\Nar{\Goutput}} \n {} \n {\Sar{\kappa}}
      \nn
      {\RELid_{j}} \n {\Ear{\iota}} \n {\RELid_{j + k}}
      \nn
      {} \n {} \n {\Sar{\Ginput}}
      \nn
      {} \n {} \n {A_2}
\diag
\end{minipage}}     
%}}}     
\hfill
$\Bkar{\displaystyle\Xi'}$
\hfill
%{{{ G -> (k, A_1, A_2)
\fbox{\begin{minipage}[c]{0.23\columnwidth}
\DIAGV{60}
      {} \n {} \n {A_1}
      \nn
      {} \n {} \n {\Nar{\Goutput}}
      \nn
      {\RELid_{i}} \n {\Ear{\iota}} \n {\RELid_{i + k}}
      \nn
      {} \n {} \n {\Nar{\kappa}}
      \nn
      {} \n {} \n {\RELid_{k}}   
      \nn
      {} \n {} \n {\Sar{\kappa}}
      \nn
      {\RELid_{j}} \n {\Ear{\iota}} \n {\RELid_{j + k}}
      \nn
      {} \n {} \n {\Sar{\Ginput}}
      \nn
      {} \n {} \n {A_2}
\diag
\end{minipage}}     
%}}}     
\hfill
$\Ar{\displaystyle\Omega'}$
\hfill
%{{{ R -> (k, A_1, A_2)
\fbox{\begin{minipage}[c]{0.23\columnwidth}
\DIAGV{60}
      {} \n {} \n {A_1}
      \nn
      {} \n {} \n {\Nar{\Goutput}}
      \nn
      {\RELid_{i}} \n {\Ear{\iota}} \n {\RELid_{i + k}}
      \nn
      {\Sar{\Ginput}} \n {} \n {\Nar{\kappa}}
      \nn
      {R} \n {} \n {\RELid_{k}}   
      \nn
      {\Nar{\Goutput}} \n {} \n {\Sar{\kappa}}
      \nn
      {\RELid_{j}} \n {\Ear{\iota}} \n {\RELid_{j + k}}
      \nn
      {} \n {} \n {\Sar{\Ginput}}
      \nn
      {} \n {} \n {A_2}
\diag
\end{minipage}}     
%}}}     
\hfill
\strut
\\[-1.8ex]
\DIAGV{67}
{\Sbiar{\Gcolimit}{}}
\n {} \n {} \n {} \n {} \n {} \n
{\Sbiar{\Gcolimit}{}}  
\n {} \n {} \n {} \n {} \n {} \n
{\Sbiar{\Gcolimit}{}} 
\nn
{A} \n {} \n {} \n {\B\Warv{\Xi}{200}} \n {} \n {} \n
{H} \n {} \n {} \n {\B\Earv{\Omega}{200}} \n {} \n {} \n {B}
\diag
%}}}
\end{minipage}
  
\bigskip
\noindent
The key observation is now that
for a redex with $\bot$ as gluing graph
and injective rule LHS $\Phi$ and injective matching $M_1$
satisfying the gluing condition,
the DPO derivation step in the category of DHG matchings
can be factored over a completely standard DPO diagram
in a category of diagrams over the
category of DHG matchings,
as indicated in the
%following nested diagram:
nested diagram above.

The double-square diagram in the middle there \textbf{is a double pushout}
in the category of diagrams over the category of DHG matchings
with \emph{rigid diagram homomorphisms},
which we define to be diagram homomorphisms that have only identity
morphisms as components,
or, in other words, that are node- and edge-label preserving graph
homomorphisms between the underlying node- and edge-labelled graphs
of the diagrams.

A key ingredient for this factoring to work is the restriction of the gluing
graph to a ``pure interface'' $\bot_{i,j}$,
so that it does not need to occur ``in the place of $L$''.
It is crucial that this place is empty in the gluing and host diagrams,
since otherwise we would not have rigid diagram homomorphisms horizontally.

As a result, since \textbf{the $\ensuremath{\Conid{COLIMIT}}$ functor preserves pushouts},
the context decomposition carries over to the result $B$
of the original DPO rewrite step, and we have:
$$
B \quad\cong\quad A_1\ \fcmp\ (R\ \otimes\ \RELid_{k})\ \fcmp\ A_2
$$
All this together proves:

\medskip
\begin{The}\Thelabel{ContextPreservation}
Let a DHG homomorphism span $L \Bkar{\Phi} \bot_{i,j} \Ar{\Psi} R$
be given where $L$ and $R$ are term graphs.
If $A : \TG{m}{n}$ is a term graph,
$\ M_1 : L \tfun A\ $ is an injective term graph matching
that together with $\Phi$ satisfies the dangling condition,
and  $(k, A_1, A_2)$ is an image context for $M_1$,
then the result graph $B$ of the induced DPO in the category of DHG
matchings
is isomorphic to $\ A_1\ \fcmp\ (R\ \otimes\ \RELid_{k})\ \fcmp\ A_2\,$,
that is, the same $(k, A_1, A_2)$ is also an image context for the morphism
$\ M_2 : R \tfun B\ $ resulting from the DPO.
\qed
\end{The}

\medskip
Note that this result is independent of the choice of image context
for $M_1$.
(Unlike for \Theref{TG-DPO}, we did not need to restrict $\Phi$ to be
injective here. Injectivity of $M_1$ however is needed for the ``image
context for'' statements according to \Defref{ImageContextFor},
and ultimately for making $M_1'$ a rigid diagram homomorphism.)

\medskip
\let\RELcomp=\origRELcomp
Let us now assume a semantics to be chosen,
that is, some gs-monoidal category (e.g., $\SetCat$),
and one of its objects $\CalV$ as interpretation of $1$.
We will use ``$\RELcomp$'' as sequential composition
and ``$\times$'' as parallel (that is, monoidal) composition in the
semantics category.

For a term graph $J : \TG{m}{n}$, we write $\sem{J}_{m,n}$
for its semantics, which is a morphism from $\CalV^m$ to $\CalV^n$.
In other words, we denote the morphism component of the semantics
functor with $\sem{\_}$; since this is a gs-monoidal functor,
we have in particular
$\sem{J_1\ \fcmp\ J_2} = \sem{J_1}\ \RELcomp\ \sem{J_2}$
and
$\sem{J_1\ \otimes\ J_2} = \sem{J_1}\ \times\ \sem{J_2}$.

\smallskip
Under the assumption that the rule $L \Bkar{} \bot_{i,j} \Ar{} R$ is semantics preserving,
that is,
$\sem{L}_{i,j} = \sem{R}_{i.j}$,
we therefore easily obtain semantics preservation of the rewrite result:
$$\renewcommand{\arraystretch}{1.3}\begin{array}{rcl}
\sem{A}_{m,n}
&=&
\sem{A_1\ \fcmp\ (L\ \otimes\ \RELid_{k})\ \fcmp\ A_2}_{m,n}
\\ &=&
\sem{A_1}_{m,i+k}\ \RELcomp\ (\sem{L}_{i,j}\ \times\ \sem{\RELid_{k}}_{k,k})\ \RELcomp\ \sem{A_2}_{j+k,n}
\\ &=&
\sem{A_1}_{m,i+k}\ \RELcomp\ (\sem{R}_{i,j}\ \times\ \sem{\RELid_{k}}_{k,k})\ \RELcomp\ \sem{A_2}_{j+k,n}
\\ &=&
\sem{A_1\ \fcmp\ (R\ \otimes\ \RELid_{k})\ \fcmp\ A_2}_{m,n}
\\ &=&
\sem{B}_{m,n}
\end{array}
$$

\medskip
%\noindent
For rules with $\bot_{i,j}$ as gluing graph,
this, 
together with \Theref{ContextPreservation}, allows us to extend
\Theref{TG-DPO}
with semantics preservation:

\begin{The}
If a term graph rewrite rule formulated as a span
$\ L \Bkar{\Phi} \bot_{i,j} \Ar{\Psi} R\ $
of DHG homomorphisms with term graphs $L,\ R : \TG{i}{j}$,
and with injective $\Phi$,
is applied via an injective term graph matching $M_1$
to an application term graph $\ A : \TG{m}{n}$,
where $M_1$ together with $\Phi$ satisfies the dangling condition,
then the diagram
  
\DIAGV{65}
{L} \n {\War{\Phi}} \n {\bot_{i,j}} \n {\Ear{\Psi}} \n {R}
\nn
{\Sar{M_{\mathrm{1}}}}
\nn
{A}
\diag
can be completed to a TG-DPO
%double-pushout diagram
\DIAGV{65}
{L} \n {\War{\Phi}} \n {\bot_{i,j}} \n {\Ear{\Psi}} \n {R}
\nn
{\Sar{M_{\mathrm{1}}}} \n {} \n {\Sar{X}} \n {} \n {\saR{M_{\mathrm{2}}}}
\nn
{A} \n {\B\War{\Xi}} \n {H} \n {\B\Ear{\Omega}} \n {B}
\diag

\smallskip
\noindent
%in the category of DHG matchings, where all horizontal arrows are
%term graph homomorphisms,
and for any gs-monoidal semantics functor
$\sem{\_}$
for which the rule is semantics-preserving, that is,
\hbox{$\sem{L}_{i,j} = \sem{R}_{i.j}$,}
the resulting TG-DPO rewrite is also semantics-preserving, that is, 
$\sem{A}_{m,n} = \sem{B}_{m,n}$.  
\qed
\end{The}
%}}}

\ignore{
%{{{ \section{GS-Monoidal Categories of Jungle Term Graphs}
\section{GS-Monoidal Categories of Jungle Term Graphs}

\edcomm{WK}{Taken from \cite{Kahl-Anand-Carette-2005}.}
%{{{ Term graphs
Term graphs are usually represented
by graphs where nodes are labelled with function symbols
and edges connect function calls with their arguments
\cite{Sleep-Plasmeijer-vanEekelen-1993}.
An alternative representation was
introduced with the name of \emph{jungle}
by Hoffmann and Plump \cite{Hoffmann-Plump-1988}
for the purpose of efficient implementation of term rewriting systems
(it is called ``term graph'' in \cite{Plump-1999}).

A \emph{jungle} is a directed hypergraph
where nodes are only labelled with type information (if applicable),
function names are hyperedge labels,
each hyperedge has a sequence of input tentacles and
exactly one output tentacle,
and for each node,
there is at most one hyperedge
that has its output tentacle incident with that node.

For representing our declarative assembly code fragments,
we use a generalisation of the jungle concept,
corresponding to \Stefanescu{}'s ``flow graphs'' \cite{Stefanescu-2000}:
%}}}

%{{{ {Def}\Deflabel{CodeGraph}
\begin{Def}\Deflabel{CodeGraph}
A \emph{code graph} $G = (\Node,\Edge,\cgIn,\cgOut,\cgSrc,\cgTrg,\cgELab)$
over an edge label set $\ELab$
consists of
\begin{itemize}
\item a set $\Node$ of \emph{nodes}
 and a set $\Edge$ of \emph{hyperedges} (or \emph{edges}),
\item two node sequences $\cgIn, \cgOut : \Node^*$
 containing the \emph{input nodes} and \emph{output nodes} of the code graph,
\item two functions $\cgSrc, \cgTrg : \Edge \tfun \Node^*$
 assigning each hyperedge the sequence of its \emph{source nodes}
 and \emph{target nodes}
 respectively, and

%%%\item a function $\cgSrc : \Edge \tfun \Node^*$
%%% assigning each hyperedge the sequence of its \emph{source nodes},
%%%\item a function $\cgTrg : \Edge \tfun \Node^*$
%%% assigning each hyperedge the sequence of its \emph{target nodes}, and

\item a function $\cgELab : \Edge \tfun \ELab$
 assigning each hyperedge its \emph{edge label},
 where the label has to be compatible with the numbers of source and
 target nodes of the edge.
\qed
\end{itemize}
\end{Def}
%}}}

We now summarise the theory of our code graphs,
which is essentially a reformulation of
\Stefanescu{}'s data-flow network algebra,
in the language of category theory.
In particular, we use the gs-monoidal categories
proposed by Corradini and Gadducci
for modelling acyclic term graphs \cite{Corradini-Gadducci-1999-APTG}.

%{{{ category ... gs-monoidal category
The following definition serves mainly to introduce our notation:

%{{{ {Def}{category}
\begin{Def}
A \emph{category} ${\catC}$
is a tuple $(\Obj, \Mor, \src, \trg, \RELid, \RELcomp)$
with the following constituents:
\begin{itemize}
\item $\Obj$ is a collection of \emph{objects}.
\item $\Mor$ is a collection of \emph{arrows} or \emph{morphisms}.
\item $\src$ (resp.~$\trg$) maps each morphism
  to its source (resp.~target) object.

We write ``$f : \objA \tfun \objB$''
for ``$f \in \Mor \land \src(f) = \objA \land \trg(f) = \objB$''.
The collection of all morphisms $f$ of category $\catC$
with $f : \objA \tfun \objB$
is denoted as $\CThom{\catC}{\objA}{\objB}$ and also called a \emph{homset}.

\item ``$\RELcomp$'' is the binary \emph{composition} operator,
  and composition of two morphisms $f : \objA \tfun \objB$ and
  $g : \objB' \tfun \objC$ is defined iff $\objB = \objB'$,
  and then $(f \RELcomp g) : \objA \tfun \objC$;
  composition is associative.
\item $\RELid$ associates with every object $\objA$ a morphism $\RELid_{\objA}$
  which is both a right and left unit for composition.
\QED
\end{itemize}
\end{Def}
%}}}

\ignore{
%{{{ {Not}{category}
\begin{Not}
For a \emph{category}
${\catC} = (\Obj_{\catC}, \Mor_{\catC}, \src, \trg, \RELid, \RELcomp)$,
we write $f : \objA \tfun \objB$
instead of $\src(f) = \objA \land \trg(f) = \objB$;
composition of %two morphisms
$f : \objA \tfun \objB$ and
  $g : \objB \tfun \objC$ is written $f \RELcomp g$;
and the identity for object $\objA$ is $\RELid_{\objA}$.
\qed
\end{Not}
%}}}
}%ignore

The objects of the untyped code graph category
over a set of edge labels $\ELab$
are natural numbers;
in the typed case we would have sequences of types.
A morphism from $m$ to $n$ is a code graph with $m$ input nodes
and $n$ output nodes
(more precisely, it is an isomorphism class of code graphs,
since node and edge identities do not matter).
Composition $F \RELcomp G$ ``glues'' together
the output nodes of $F$ with the respective input nodes of $G$.
The identity on $n$ consists only of $n$ input nodes which are also,
in the same sequence, output nodes, and no edges.

A \emph{primitive} code graph
is a code graph that corresponds to a single operation,
i.e., a code graph with a single edge
where each node is the target of exactly one tentacle,
and the target node sequence of the edge coincides with the
output node sequence of the graph,
and the source sequence with the input sequence.

%{{{ {Def}[\cite{MacLane-1971}]\Deflabel{ssmc}
\begin{Def}\Deflabel{ssmc}
A \emph{symmetric strict monoidal category} %\cite{MacLane-1971}
$\catC = ( \catCo, \otimes, \munit, \exch )$
consists of a category $\catCo$,
a strictly associative monoidal bifunctor $\otimes$
with $\munit$ as its strict unit,
%%% i.e.,
%%% for objects $\objA$, $\objB$, and $\objC$ we have the monoid laws
%%% \BD
%%%     \objA \otimes (\objB \otimes \objC)
%%%   \sepA{=}
%%%     (\objA \otimes \objB) \otimes \objC
%%% \qquad\mbox{and}\qquad
%%%    \objA \otimes \triv
%%%   \sepA{=}
%%%    \objA
%%%   =
%%%    \triv \otimes \objA
%%% \enskip,
%%% \ED
%%% and for arrows
%%% we have
%%% \begin{itemize}
%%% \item \emph{associativity}: \BM
%%%     F \otimes (G \otimes H) = (F \otimes G) \otimes H
%%%   \EM{},
%%% \item \emph{functoriality}: \BM
%%%     (F \otimes G) \RELcomp (H \otimes J)
%%%   =
%%%     (F \RELcomp H) \otimes (G \RELcomp J)
%%%   \EM{},
%%% \item \emph{preservation of identities}: \BM
%%%     \RELid_{\objA} \otimes \RELid_{\objB} = \RELid_{\objA \otimes \objB}
%%%   \EM{}, and
%%% \item \emph{monoidal unit on arrows}: \BM
%%%     \RELid_{\triv} \otimes F = F
%%%   \EM{} and \BM
%%%     F \otimes \RELid_{\triv} = F
%%%   \EM{}.
%%% \end{itemize}
%%%
%%% \noindent
%%% Furthermore,
and a transformation $\exch$
that associates with every two objects $\objA$ and $\objB$
an arrow $\exch_{\objA,\objB} : \objA \otimes \objB \tfun \objB \otimes \objA$
with:
\BCM
\begin{array}[b]{rcl@{\hskip1em}rcl}
    (F \otimes G) \RELcomp \exch_{\objC,\objD} &=&
    \exch_{\objA,\objB} \RELcomp (G \otimes F)
\enskip,&
    \exch_{\objA,\objB} \RELcomp \exch_{\objB,\objA} &=&
    \RELid_{\objA} \otimes \RELid_{\objB}
\enskip,\\[.3ex]
    \exch_{\objA\otimes\objB,\objC} &=&
    (\RELid_{\objA} \otimes \exch_{\objB,\objC}) \RELcomp
    (\exch_{\objA,\objC} \otimes \RELid_{\objB})
\enskip,&
    \exch_{\munit,\munit} &=& \RELid_{\munit}
\enskip.
\ECMAQ
\unskip
\end{Def}
%}}}

\noindent
For code graphs, $\munit$ is the number 0 and
$\otimes$ on objects is addition.
On morphisms, $\otimes$ forms the disjoint union of code graphs,
concatenating the input and output node sequences.
$\exch_{m,n}$ differs from $\RELid_{m + n}$
only in the fact that the two parts of the output node sequence are swapped.

\ignore{
%{{{ {Def}[gs-monoidal categories]\Deflabel{gs-monoidal}
\begin{Def}\Deflabel{gs-monoidal}
$\catC = ( \catCo, \otimes, \munit, \exch, \Nabla, !  )$
is a {\em strict gs-monoidal category} iff
\begin{itemize}
\item $( \catCo, \otimes, \munit, \exch )$
  is a symmetric strict monoidal category, and
\item $!$ associates with every object $\objA$ of $\catCo$
  an arrow $!_{\objA} : \objA \tfun \munit$, and
\item $\Nabla$ associates with every object $\objA$ of $\catCo$
  an arrow $\NablaU{\objA} : \objA \tfun \objA \otimes \objA$,
\end{itemize}
such that $\RELid_{\munit} = !_{\munit} = \NablaU{\munit}$,
and the following axioms hold:

\BCM
\def\arraystretch{1.3}
\begin{array}[b]{l}
    \NablaU{\objA} \RELcomp (\RELid_{\objA} \otimes \NablaU{\objA})
  \sepA{=}
    \NablaU{\objA} \RELcomp (\NablaU{\objA} \otimes \RELid_{\objA})
\qquad
    \NablaU{\objA} \RELcomp \exchU{\objA,\objA}
  \sepA{=}
    \NablaU{\objA}
\qquad
    \NablaU{\objA} \RELcomp (\RELid_{\objA} \otimes !_{\objA})
  \sepA{=}
    \RELid_{\objA}
\\
    \NablaU{\objA \otimes \objB} \RELcomp
    (\RELid_{\objA} \otimes \exchU{\objB,\objA} \otimes \RELid_{\objB})
  \sepA{=}
    \NablaU{\objA} \otimes \NablaU{\objB}
\qquad
\qquad
    !_{\objA \otimes \objB}
  \sepA{=}
    !_{\objA} \otimes !_{\objB}
\ECMAQ
\end{Def}
%}}}
}%ignore

%{{{ {Def}[g-monoidal categories]\Deflabel{g-monoidal}
\begin{Def}\Deflabel{g-monoidal}
$\catC = ( \catCo, \otimes, \munit, \exch,  !  )$
is a {\em strict g-monoidal category} iff
\begin{itemize}
\item $( \catCo, \otimes, \munit, \exch )$
  is a symmetric strict monoidal category, and
\item $!$ associates with every object $\objA$ of $\catCo$
  an arrow $!_{\objA} : \objA \tfun \munit$,
\end{itemize}
such that $\RELid_{\munit} = !_{\munit}$,
and \emph{monoidality of termination} holds:
\BM
    !_{\objA \otimes \objB}
  \sepA{=}
    !_{\objA} \otimes !_{\objB}
\EMQ
\end{Def}
%}}}

\noindent
For code graphs,
$!_n$ differs from $\RELid_n$
only in the fact that the output node sequence is empty.
The ``g'' of ``g-monoidal'' stands for ``garbage'':
all edges of code graph $G : m \tfun n$
are backward-garbage in $G \RELcomp !_n$.

Note that $!_n$ itself is garbage free, coherent, and lean, and
therefore solid and even executable.

%{{{ {Def}[s-monoidal categories]\Deflabel{s-monoidal}
\begin{Def}\Deflabel{s-monoidal}
$\catC = ( \catCo, \otimes, \munit, \exch, \Nabla  )$
is a {\em strict s-monoidal category} $\catC$ iff
\begin{itemize}
\item $( \catCo, \otimes, \munit, \exch )$
  is a symmetric strict monoidal category, and
\item $\Nabla$ associates with every object $\objA$ of $\catCo$
  an arrow $\NablaU{\objA} : \objA \tfun \objA \otimes \objA$,
\end{itemize}
such that $\RELid_{\munit} = \NablaU{\munit}$,
and the \emph{coherence} axioms
\begin{itemize}
\item  \emph{associativity of duplication}:
\BM
    \NablaU{\objA} \RELcomp (\RELid_{\objA} \otimes \NablaU{\objA})
  \sepA{=}
    \NablaU{\objA} \RELcomp (\NablaU{\objA} \otimes \RELid_{\objA})
\EM,
\item \emph{commutativity of duplication}:
\BM
    \NablaU{\objA} \RELcomp \exchU{\objA,\objA}
  \sepA{=}
    \NablaU{\objA}
\EM
\end{itemize}
and the \emph{monoidality} axiom
\begin{itemize}
\item \emph{monoidality of duplication}:
\BM
    \NablaU{\objA \otimes \objB} \RELcomp
    (\RELid_{\objA} \otimes \exchU{\objB,\objA} \otimes \RELid_{\objB})
  \sepA{=}
    \NablaU{\objA} \otimes \NablaU{\objB}
\EM
\end{itemize}
are satisfied.
\qed
\end{Def}
%}}}

For code graphs,
$\Nabla_{n}$ differs from $\RELid_{n}$
only in the fact that the output node sequence is
\ignore{
and in
$\Nabla_{\objA}$, the output node sequence is
}%ignore
the concatenation of the input node sequence with itself.
The ``s'' of ``s-monoidal'' stands for ``sharing:
every input of $\Nabla_k \RELcomp (F \otimes G)$ is shared by $F : k \tfun m$
and $G : k \tfun n$.

%{{{ {Def}[gs-monoidal categories]\Deflabel{gs-monoidal}
\begin{Def}\Deflabel{gs-monoidal}
$\catC = ( \catCo, \otimes, \munit, \exch, \Nabla, !  )$
is a {\em strict gs-monoidal category} iff
\begin{itemize}
\item $( \catCo, \otimes, \munit, \exch, ! )$
  is a strict g-monoidal category, and
\item $( \catCo, \otimes, \munit, \exch, \Nabla )$
  is a strict s-monoidal category,
\end{itemize}
such that the \emph{coherence} axiom
\begin{itemize}
\item \emph{right-inverse of duplication} holds:
\BM
    \NablaU{\objA} \RELcomp (\RELid_{\objA} \otimes !_{\objA})
  \sepA{=}
    \RELid_{\objA}
\EMQ
\end{itemize}
%
%{{{ \ignore{gs-monoidal functor}
\ignore{
A {\em gs-monoidal functor}
$( F, \phi, \phi_e ) : \catC \rightarrow \catC'$ is a
symmetric monoidal functor (that is, a functor $F$ equipped with two natural
isomorphisms $\phi_e: F(e) \rightarrow e'$ and \
$\phi: F(\objA \otimes \objB) \rightarrow F(a) \otimes' F(b)$)
such that $F(!_{\objA}) ; \phi_e = !_{F(a)}$ and $F(\NablaU{\objA}) ; \phi =
\NablaU{F(a)}$;
it is {\em strict} if $\phi$ and $\phi_e$ are identities.
The category of small strict gs-monoidal categories and their strict
functors is denoted by {\bf GSM-Cat}.
}%ignore
%}}}
%
\end{Def}
%}}}
%}}}

\noindent
%{{{ gs-monoidal category, but not a \emph{cartesian} category
Code graphs (and term graphs) over a fixed edge label set
form a gs-monoidal category, but not a \emph{cartesian} category,
where in addition $!$ and $\Nabla$ are \emph{natural} transformations,
i.e., for all $F : \objA \tfun \objB$
we have
$F \RELcomp !_{\objB} = !_{\objA}$
and
\hbox{$F \RELcomp \NablaU{\objB} = \NablaU{\objA} \RELcomp (F \otimes F)$.}
\ignore{
%{{{ {Def}[cartesian categories]\Deflabel{cartesian}
\begin{Def}\Deflabel{cartesian}
A {\em strict cartesian category} $\catC$
is a strict gs-monoidal category
$( \catCo, \otimes, \munit, \exch, \Nabla, !  )$,
where
\begin{itemize}
\item $!$ is a natural transformation from the identity functor to the
  constant-$\munit$ functor,
  i.e., $F \RELcomp !_{\objB} = !_{\objA}$ for all $F : \objA \tfun \objB$,
  and
\item $\Nabla$ is a natural transformation from the identity functor to
  $\otimes$,
  i.e., $F \RELcomp \NablaU{\objB} = \NablaU{\objA} \RELcomp (F \otimes F)$
  for all $F : \objA \tfun \objB$
\qed
\end{itemize}
\end{Def}
%}}}
}%ignore
%
%\noindent
To see how these naturality conditions are violated,
the five jungles in the following drawing
can be obtained as, in this sequence,
$F : 1 \tfun 1$,
$!_1$,
$F \RELcomp !_1$,
$F \RELcomp \Nabla_1$,
and
$\Nabla_1 \RELcomp (F \otimes F)$:

\phantom{.}\hfill
\CGpic{gF}\hfill % 35 35 95 325
\CGpicO{bang}{bb=35 -143 95 147}\hfill  % 35 35 95 147
\CGpicO{gFbang}{bb=35 -45 95 245}\hfill % 35 35 95 245
\CGpic{gFdup}\hfill
\CGpic{dup_gF}\hfill
%\CGpic{gQ}\hfill
%\CGpic{gQbang}\hfill
\phantom{.}

\smallskip
\noindent
(It is easy to see that we obtain naturality of termination
if we consider
equivalence classes of code graphs up to garbage collection.)
%}}}
%}}}
}%ignore

\ignore{
%{{{ \section{Modelling Term Graphs using ``Dependent Objects''}\sectlabel{TG}
\section{Modelling Term Graphs using ``Dependent Objects''}\sectlabel{TG}

\edcomm{WK}{Refer to \citep{Kahl-2011_AgdaTG}.}

\edcomm{WK}{The following material is unchanged from the MSFP 2018 submission.}

\noindent
Term graphs are ``expression trees with sharing (and possibly cycles)'';
following \citet{Hoffmann-Plump-1991} and
\citet{Corradini-Rossi-1991},
we consider term graphs as a kind of directed hypergraphs (``jungles''),
where the operation labels are attached to the hyperedges.
For simplicity, we will not formalise any requirement of acyclicity,
nor of bijective correspondence between hyperedges and non-variable nodes.

However, we do want to enforce that each edge $e$ has an arity,
and that the label of $e$ is taken from a label set for that arity,
and that the arity is the number of argument nodes (``source nodes'') for this edge.

Our first formalisation of this uses a dependent sum construction:
In general, the dependent sum type $(\sum x : X \ \bullet\ T\ x)$
is the type of pairs $(x, y)$ where $x$ is of type $X$ and $y$ is of
type $T\ x$, which means that the \emph{type} of $y$ depends on
the \emph{value} of $x$.

The first function symbol $\Fct{trg}$ of $\mathsf{sigTG_1}$
maps each edge to its result (``target'') node;
the second function symbol maps each edge to a dependently-typed
triple $(n, q, a)$ consisting of the arity $n$, the edge label $q$ of
type $\ELab\ n$, and the array $a$ of argument nodes, of type
$\ensuremath{\Conid{Vec}}\ \Sort{N}\ n$ (which we expect to be implemented in structure
$G$ by $\ensuremath{\Conid{Array}}\ \Sort{N}_G\ n$):
%
%{{{ sigTG_1  \sum
\BD
    \mathsf{sigTG_1}
  \defeq
    \begin{array}[t]{ll}
    \langle & \mbox{\textbf{sorts: }} \Sort{N}, \Sort{E}
    \\
    {} & \mbox{\textbf{ops: }} \begin{array}[t]{l}
          \Fct{trg} : \Sort{E} \tfun \Sort{N}
       \\ \Fct{eInfo} : \Sort{E} \tfun \sum \ n : \nat\ \bullet\ \ELab\ n \times \ensuremath{\Conid{Vec}}\ \Sort{N}\ n
  \kern1em\rangle
       \end{array}
%    \\
%    \rangle
    \end{array}
    \ED
%}}}

\medbreak
\noindent
A nicer formulation is possible using \emph{dependent function types}
--- in general, the type ``$(x : X) \tfun T\ x$'' is the type of
functions
that map each argument $x$ of type $X$ to a result of type $T\ x$,
that is, not only the value of the result depends on the value of the
argument,
but also the \emph{type} of the result depends on the \emph{value} of
the argument.

Using such dependent function types,
we can split the triple type of $\Fct{eInfo}$ similar to the splitting
mentioned at the beginning of this section, which however split a
type of non-dependent pairs.
Here, while introducing
separate function symbols for arity, label, and source node
array of edges,
we also have to turn the types of the latter two function symbols
into dependent function types:
%
%{{{ sigTG_2
\BD
    \mathsf{sigTG_2}
  \defeq
    \begin{array}[t]{ll}
    \langle & \mbox{\textbf{sorts: }} \Sort{N}, \Sort{E}
    \\
    {} & \mbox{\textbf{ops: }} \begin{array}[t]{l}
          \Fct{trg} : \Sort{E} \tfun \Sort{N}
       \\ \Fct{arity} : \Sort{E} \tfun \nat
       \\ \Fct{lab} : (e : \Sort{E}) \tfun \ELab \ (\Fct{arity}\ e)
       \\ \Fct{src} : (e : \Sort{E}) \tfun \ensuremath{\Conid{Vec}}\ \Sort{N}\ (\Fct{arity}\ e)
  \kern1em\rangle
       \end{array}
%    \\
%    \rangle
    \end{array}
    \ED
%}}}

\noindent
For implementation in the setting of \sectref{VecSG},
this setup does not present significant problems:
For a $\mathsf{sigTG_2}$-structure $G$,
the interpretation $\Fct{lab}_G$ can be implemented by an array of type
$$
\ensuremath{\Conid{Array}}\ (\sum n : \nat \ \bullet\ \ELab \ n)\ (\#\ \Sort{E}_G)
\enskip,
$$
where $\#\ \Sort{E}_G$ stands for the cardinality of the carrier set
of $\Sort{E}$ in $G$.
Such structures will then be subject to the datatype invariant
that the images of $\Fct{lab}$ conform with those
of $\Fct{arity}$, and analogous for $\Fct{src}$.
(Presence of this datatype invariant makes it possible
to implement $(\sum n : \nat \ \bullet\ \ELab \ n)$ as an untagged
union type, which makes the correspondence with
$\mathsf{sigTG_1}$-structures more direct.)
%}}}
}%ignore

%{{{ \section{Conclusion and Outlook}
\section{Conclusion and Outlook}

By considering a straight-forward adaptation of the DPO approach to
term graph rewriting,
we obtained an easily-understandable concept of rule application.
By lifting this adapted DPO into a standard DPO of diagrams,
we have been able to transfer the context decomposition
from the left-hand side to the right-hand side, obviating the need to consider
any semantics for general DHGs such as $\bot_{i,j}$.
As result, we obtained a semantics preservation theorem
that will be an important tool
in the generation of verified code optimisation tools
employing rule-based transformation of data-flow graphs,
as outlined for example in \cite{Kahl-2014_Mouldable}.

%\edcomm{R2}{If possible I would be curious to hear more about the use of the presented
%  approach as mentioned in Conclusion and Outlook.}

We originally started in
\cite{Kahl-2011_AgdaTG}
to formalise term graphs essentially as defined in
\sectref{gsMonCat} in the dependently-typed programming language
and proof assistant Agda \cite{Norell-2007}.
The current status of this project
\cite{Kahl-2017_RATH-Agda-2.2,Zhao-2018_TGR1}
includes term graph decomposition
and a proof for its correctness,
which essentially constitutes a machine-checked proof of the
result of Corradini and Gadducci
\cite{Corradini-Gadducci-1999-APTG}
that term graphs form a free gs-monoidal category.
As next steps, we plan to extend this development
to cover also the results of the current paper,
that is, definedness and semantics preservation of TG-DPO rewriting
steps, and then to use this as a verified implementation
of semantics-preserving term graph rewriting.

%\citep{Kahl-2011_AgdaTG}
%\unfinished

%}}}

%\footnotesize
%\let\url=\texttt
%\bibliographystyle{relmics-alpha-lncs}
%\bibliographystyle{eceasst}
%\bibliographystyle{splncs03LNCS}
%\bibliographystyle{abbrvnatLNCS} % (natbib)
%\bibliographystyle{eptcs}
%\bibliographystyle{eptcsaplpha}
%\bibliographystyle{eptcsini}
\bibliographystyle{eptcsalphaini}
%%% \def\bibsection{%
%%%    \section*{\refname}%
%%% %   \section*{\refname\@mkboth{\MakeUppercase{\refname}}{\MakeUppercase{\refname}}}%
%%%   }%
%\bibliography{strings,ref,crossrefs}
%\bibliography{AgdaGT1}

\input TGR.bbl
\end{document}

%% file: diagram-2.2.tex
\newcommand{\tlowername}[2]%
{$\stackrel{\makebox[1pt]{#1}}%
{\begin{picture}(0,0)%
\put(0,0){\makebox(0,6)[t]{\makebox[1pt]{$#2$}}}%
\end{picture}}$}%

\newcommand{\Tcase}[2]{\makebox[23pt]%
{\raisebox{2.5pt}{$\stackrel{#2}{#1{20}}$}}}%

\newcommand{\AR}[1]%
{\begin{picture}(#1,0)%
\put(0,0){\vector(1,0){#1}}%
\end{picture}}%

\newcommand{\DOTAR}[1]%
{\NUMBEROFDOTS=#1%
\divide\NUMBEROFDOTS by 3%
\begin{picture}(#1,0)%
\multiput(0,0)(3,0){\NUMBEROFDOTS}{\circle*{1}}%
\put(#1,0){\vector(1,0){0}}%
\end{picture}}%

\newcommand{\MONO}[1]%
{\begin{picture}(#1,0)%
\put(0,0){\vector(1,0){#1}}%
\put(2,-2){\line(0,1){4}}%
\end{picture}}%

\newcommand{\EPI}[1]%
{\begin{picture}(#1,0)(-#1,0)%
\put(-#1,0){\vector(1,0){#1}}%
\put(-6,-2){\line(0,1){4}}%
\end{picture}}%

\newcommand{\BIMO}[1]%
{\begin{picture}(#1,0)(-#1,0)%
\put(-#1,0){\vector(1,0){#1}}%
\put(-6,-2){\line(0,1){4}}%
\put(-#1,-2){\hspace{2pt}\line(0,1){4}}%
\end{picture}}%

\newcommand{\BIAR}[1]%
{\begin{picture}(#1,4)%
\put(0,0){\vector(1,0){#1}}%
\put(0,4){\vector(1,0){#1}}%
\end{picture}}%

\newcommand{\EQL}[1]%
{\begin{picture}(#1,0)%
\put(0,1){\line(1,0){#1}}%
\put(0,-1){\line(1,0){#1}}%
\end{picture}}%

\newcommand{\ADJAR}[1]%
{\begin{picture}(#1,4)%
\put(0,0){\vector(1,0){#1}}%
\put(#1,4){\vector(-1,0){#1}}%
\end{picture}}%

\newcommand{\Ar}[1]{\Tcase{\AR}{#1}}%

\newcommand{\BKAR}[1]%
{\begin{picture}(#1,0)%
\put(#1,0){\vector(-1,0){#1}}%
\end{picture}}%

\newcommand{\BKDOTAR}[1]%
{\NUMBEROFDOTS=#1%
\divide\NUMBEROFDOTS by 3%
\begin{picture}(#1,0)%
\multiput(#1,0)(-3,0){\NUMBEROFDOTS}{\circle*{1}}%
\put(0,0){\vector(-1,0){0}}%
\end{picture}}%

\newcommand{\BKMONO}[1]%
{\begin{picture}(#1,0)(-#1,0)%
\put(0,0){\vector(-1,0){#1}}%
\put(-2,-2){\line(0,1){4}}%
\end{picture}}%

\newcommand{\BKEPI}[1]%
{\begin{picture}(#1,0)%
\put(#1,0){\vector(-1,0){#1}}%
\put(6,-2){\line(0,1){4}}%
\end{picture}}%

\newcommand{\BKBIMO}[1]%
{\begin{picture}(#1,0)%
\put(#1,0){\vector(-1,0){#1}}%
\put(6,-2){\line(0,1){4}}%
\put(#1,-2){\hspace{-2pt}\line(0,1){4}}%
\end{picture}}%

\newcommand{\BKBIAR}[1]%
{\begin{picture}(#1,4)%
\put(#1,0){\vector(-1,0){#1}}%
\put(#1,4){\vector(-1,0){#1}}%
\end{picture}}%

\newcommand{\BKADJAR}[1]%
{\begin{picture}(#1,4)%
\put(0,4){\vector(1,0){#1}}%
\put(#1,0){\vector(-1,0){#1}}%
\end{picture}}%

\newcommand{\Bkar}[1]{\Tcase{\BKAR}{#1}}%

\newcommand{\lowername}[2]%
{$\stackrel{\makebox[1pt]{#1}}%
{\begin{picture}(0,0)%
\truex{600}%
\put(0,-600){\makebox(0,\value{x})[t]{\makebox[1pt]{$#2$}}}%
\end{picture}}$}%

\newcommand{\hcase}[2]%
{\makebox[0pt]%
{\raisebox{-1pt}[0pt][0pt]{#1{#2}}}}%

\newcommand{\Hcase}[3]%
{\makebox[0pt]
{\raisebox{-1pt}[0pt][0pt]%
{$\stackrel{\makebox[0pt]{$\textstyle{#2}$}}{#1{#3}}$}}}%

\newcommand{\hcasE}[3]%
{\makebox[0pt]%
{\raisebox{-2pt}[0pt][0pt]%
{\lowername{#1{#3}}{#2}}}}%

\newcommand{\hbicase}[2]%
{\makebox[0pt]%
{\raisebox{-2.5pt}[0pt][0pt]{#1{#2}}}}%

\newcommand{\Hbicase}[4]%
{\makebox[0pt]
{\raisebox{-10.5pt}[0pt][0pt]%
{$\stackrel{\makebox[0pt]{$\textstyle{#2}$}}%
{\mbox{\lowername{#1{#4}}{#3}}}$}}}%

\newcommand{\EAR}[1]%
{\begin{picture}(#1,0)%
\put(0,0){\vector(1,0){#1}}%
\end{picture}}%

\newcommand{\EDOTAR}[1]%
{\truex{100}\truey{300}%
\NUMBEROFDOTS=#1%
\divide\NUMBEROFDOTS by \value{y}%
\begin{picture}(#1,0)%
\multiput(0,0)(\value{y},0){\NUMBEROFDOTS}%
{\circle*{\value{x}}}%
\put(#1,0){\vector(1,0){0}}%
\end{picture}}%

\newcommand{\EMONO}[1]%
{\begin{picture}(#1,0)%
\put(0,0){\vector(1,0){#1}}%
\truex{300}\truey{600}%
\put(\value{x},-\value{x}){\line(0,1){\value{y}}}%
\end{picture}}%

\newcommand{\EEPI}[1]%
{\begin{picture}(#1,0)(-#1,0)%
\put(-#1,0){\vector(1,0){#1}}%
\truex{300}\truey{600}\truez{800}%
\put(-\value{z},-\value{x}){\line(0,1){\value{y}}}%
\end{picture}}%

\newcommand{\EBIMO}[1]%
{\begin{picture}(#1,0)(-#1,0)%
\put(-#1,0){\vector(1,0){#1}}%
\truex{300}\truey{600}\truez{800}%
\put(-\value{z},-\value{x}){\line(0,1){\value{y}}}%
\put(-#1,-\value{x}){\hspace{3pt}\line(0,1){\value{y}}}%
\end{picture}}%

\newcommand{\EBIAR}[1]%
{\truex{400}%
\begin{picture}(#1,\value{x})%
\put(0,0){\vector(1,0){#1}}%
\put(0,\value{x}){\vector(1,0){#1}}%
\end{picture}}%

\newcommand{\EEQL}[1]%
{\begin{picture}(#1,0)%
\truex{200}%
\put(0,\value{x}){\line(1,0){#1}}%
\put(0,0){\line(1,0){#1}}%
\end{picture}}%

\newcommand{\EADJAR}[1]%
{\truex{400}%
\begin{picture}(#1,\value{x})%
\put(0,0){\vector(1,0){#1}}%
\put(#1,\value{x}){\vector(-1,0){#1}}%
\end{picture}}%

\newcommand{\ear}%
{\hspace{\SOURCE\unitlength}%
\hcase{\EAR}{\ARROWLENGTH}}%

\newcommand{\Earv}[2]{\Hcase{\EAR}{#1}{#200}}%

\newcommand{\Ear}[1]%
{\hspace{\SOURCE\unitlength}%
\Hcase{\EAR}{#1}{\ARROWLENGTH}}%

\newcommand{\eaR}[1]%
{\hspace{\SOURCE\unitlength}%
\hcasE{\EAR}{#1}{\ARROWLENGTH}}%

\newcommand{\edotar}%
{\hspace{\SOURCE\unitlength}%
\hcase{\EDOTAR}{\ARROWLENGTH}}%

\newcommand{\Edotar}[1]%
{\hspace{\SOURCE\unitlength}%
\Hcase{\EDOTAR}{#1}{\ARROWLENGTH}}%

\newcommand{\edotaR}[1]%
{\hspace{\SOURCE\unitlength}%
\hcasE{\EDOTAR}{#1}{\ARROWLENGTH}}%

\newcommand{\emono}%
{\hspace{\SOURCE\unitlength}%
\hcase{\EMONO}{\ARROWLENGTH}}%

\newcommand{\Emono}[1]%
{\hspace{\SOURCE\unitlength}%
\Hcase{\EMONO}{#1}{\ARROWLENGTH}}%

\newcommand{\emonO}[1]%
{\hspace{\SOURCE\unitlength}%
\hcasE{\EMONO}{#1}{\ARROWLENGTH}}%

\newcommand{\eepi}%
{\hspace{\SOURCE\unitlength}%
\hcase{\EEPI}{\ARROWLENGTH}}%

\newcommand{\Eepi}[1]%
{\hspace{\SOURCE\unitlength}%
\Hcase{\EEPI}{#1}{\ARROWLENGTH}}%

\newcommand{\eepI}[1]%
{\hspace{\SOURCE\unitlength}%
\hcasE{\EEPI}{#1}{\ARROWLENGTH}}%

\newcommand{\ebimo}%
{\hspace{\SOURCE\unitlength}%
\hcase{\EBIMO}{\ARROWLENGTH}}%

\newcommand{\Ebimo}[1]%
{\hspace{\SOURCE\unitlength}%
\Hcase{\EBIMO}{#1}{\ARROWLENGTH}}%

\newcommand{\ebimO}[1]%
{\hspace{\SOURCE\unitlength}%
\hcasE{\EBIMO}{#1}{\ARROWLENGTH}}%

\newcommand{\eiso}%
{\hspace{\SOURCE\unitlength}%
\Hcase{\EAR}{\cong}{\ARROWLENGTH}}%

\newcommand{\Eiso}[1]%
{\hspace{\SOURCE\unitlength}%
\Hcase{\EAR}{\cong#1}{\ARROWLENGTH}}%

\newcommand{\eisO}[1]%
{\hspace{\SOURCE\unitlength}%
\hcasE{\EAR}{\cong#1}{\ARROWLENGTH}}%

\newcommand{\ebiar}%
{\hspace{\SOURCE\unitlength}%
\hbicase{\EBIAR}{\ARROWLENGTH}}%

\newcommand{\Ebiar}[2]%
{\hspace{\SOURCE\unitlength}%
\Hbicase{\EBIAR}{#1}{#2}{\ARROWLENGTH}}%

\newcommand{\eeql}%
{\hspace{\SOURCE\unitlength}%
\hbicase{\EEQL}{\ARROWLENGTH}}%

\newcommand{\eadjar}%
{\hspace{\SOURCE\unitlength}%
\hbicase{\EADJAR}{\ARROWLENGTH}}%

\newcommand{\Eadjar}[2]%
{\hspace{\SOURCE\unitlength}%
\Hbicase{\EADJAR}{#1}{#2}{\ARROWLENGTH}}%

\newcommand{\WAR}[1]%
{\begin{picture}(#1,0)%
\put(#1,0){\vector(-1,0){#1}}%
\end{picture}}%

\newcommand{\WDOTAR}[1]%
{\truex{100}\truey{300}%
\NUMBEROFDOTS=#1%
\divide\NUMBEROFDOTS by \value{y}%
\begin{picture}(#1,0)%
\multiput(#1,0)(-\value{y},0){\NUMBEROFDOTS}%
{\circle*{\value{x}}}%
\put(0,0){\vector(-1,0){0}}%
\end{picture}}%

\newcommand{\WMONO}[1]%
{\begin{picture}(#1,0)(-#1,0)%
\put(0,0){\vector(-1,0){#1}}%
\truex{300}\truey{600}%
\put(-\value{x},-\value{x}){\line(0,1){\value{y}}}%
\end{picture}}%

\newcommand{\WEPI}[1]%
{\begin{picture}(#1,0)%
\put(#1,0){\vector(-1,0){#1}}%
\truex{300}\truey{600}\truez{800}%
\put(\value{z},-\value{x}){\line(0,1){\value{y}}}%
\end{picture}}%

\newcommand{\WBIMO}[1]%
{\begin{picture}(#1,0)%
\put(#1,0){\vector(-1,0){#1}}%
\truex{300}\truey{600}\truez{800}%
\put(\value{z},-\value{x}){\line(0,1){\value{y}}}%
\put(#1,-\value{x}){\hspace{-3pt}\line(0,1){\value{y}}}%
\end{picture}}%

\newcommand{\WBIAR}[1]%
{\truex{400}%
\begin{picture}(#1,\value{x})%
\put(#1,0){\vector(-1,0){#1}}%
\put(#1,\value{x}){\vector(-1,0){#1}}%
\end{picture}}%

\newcommand{\WADJAR}[1]%
{\truex{400}%
\begin{picture}(#1,\value{x})%
\put(0,\value{x}){\vector(1,0){#1}}%
\put(#1,0){\vector(-1,0){#1}}%
\end{picture}}%

\newcommand{\war}%
{\hspace{\SOURCE\unitlength}%
\hcase{\WAR}{\ARROWLENGTH}}%

\newcommand{\Warv}[2]{\Hcase{\WAR}{#1}{#200}}%

\newcommand{\War}[1]%
{\hspace{\SOURCE\unitlength}%
\Hcase{\WAR}{#1}{\ARROWLENGTH}}%

\newcommand{\waR}[1]%
{\hspace{\SOURCE\unitlength}%
\hcasE{\WAR}{#1}{\ARROWLENGTH}}%

\newcommand{\wdotar}%
{\hspace{\SOURCE\unitlength}%
\hcase{\WDOTAR}{\ARROWLENGTH}}%

\newcommand{\Wdotar}[1]%
{\hspace{\SOURCE\unitlength}%
\Hcase{\WDOTAR}{#1}{\ARROWLENGTH}}%

\newcommand{\wdotaR}[1]%
{\hspace{\SOURCE\unitlength}%
\hcasE{\WDOTAR}{#1}{\ARROWLENGTH}}%

\newcommand{\wmono}%
{\hspace{\SOURCE\unitlength}%
\hcase{\WMONO}{\ARROWLENGTH}}%

\newcommand{\Wmono}[1]%
{\hspace{\SOURCE\unitlength}%
\Hcase{\WMONO}{#1}{\ARROWLENGTH}}%

\newcommand{\wmonO}[1]%
{\hspace{\SOURCE\unitlength}%
\hcasE{\WMONO}{#1}{\ARROWLENGTH}}%

\newcommand{\wepi}%
{\hspace{\SOURCE\unitlength}%
\hcase{\WEPI}{\ARROWLENGTH}}%

\newcommand{\Wepi}[1]%
{\hspace{\SOURCE\unitlength}%
\Hcase{\WEPI}{#1}{\ARROWLENGTH}}%

\newcommand{\wepI}[1]%
{\hspace{\SOURCE\unitlength}%
\hcasE{\WEPI}{#1}{\ARROWLENGTH}}%

\newcommand{\wbimo}%
{\hspace{\SOURCE\unitlength}%
\hcase{\WBIMO}{\ARROWLENGTH}}%

\newcommand{\Wbimo}[1]%
{\hspace{\SOURCE\unitlength}%
\Hcase{\WBIMO}{#1}{\ARROWLENGTH}}%

\newcommand{\wbimO}[1]%
{\hspace{\SOURCE\unitlength}%
\hcasE{\WBIMO}{#1}{\ARROWLENGTH}}%

\newcommand{\wiso}%
{\hspace{\SOURCE\unitlength}%
\Hcase{\WAR}{\cong}{\ARROWLENGTH}}%

\newcommand{\Wiso}[1]%
{\hspace{\SOURCE\unitlength}%
\Hcase{\WAR}{#1}{\ARROWLENGTH}}%

\newcommand{\wisO}[1]%
{\hspace{\SOURCE\unitlength}%
\hcasE{\WAR}{#1}{\ARROWLENGTH}}%

\newcommand{\wbiar}%
{\hspace{\SOURCE\unitlength}%
\hbicase{\WBIAR}{\ARROWLENGTH}}%

\newcommand{\Wbiar}[2]%
{\hspace{\SOURCE\unitlength}%
\Hbicase{\WBIAR}{#1}{#2}{\ARROWLENGTH}}%

\newcommand{\weql}%
{\hspace{\SOURCE\unitlength}%
\hbicase{\EEQL}{\ARROWLENGTH}}%

\newcommand{\wadjar}%
{\hspace{\SOURCE\unitlength}%
\hbicase{\WADJAR}{\ARROWLENGTH}}%

\newcommand{\Wadjar}[2]%
{\hspace{\SOURCE\unitlength}%
\Hbicase{\WADJAR}{#1}{#2}{\ARROWLENGTH}}%

\newcommand{\Vcase}[3]{\makebox[0pt]%
{\makebox[0pt][r]{\raisebox{0pt}[0pt][0pt]{${#2}\hspace{2pt}$}}}#1{#3}}%

\newcommand{\vcasE}[3]{\makebox[0pt]%
{#1{#3}\makebox[0pt][l]{\raisebox{0pt}[0pt][0pt]{\hspace{2pt}$#2$}}}}%

\newcommand{\Vbicase}[4]{\makebox[0pt]%
{\makebox[0pt][r]{\raisebox{0pt}[0pt][0pt]{$#2$\hspace{4pt}}}#1{#4}%
\makebox[0pt][l]{\raisebox{0pt}[0pt][0pt]{\hspace{5pt}$#3$}}}}%

\newcommand{\SAR}[1]%
{\begin{picture}(0,0)%
\put(0,0){\makebox(0,0)%
{\begin{picture}(0,#1)%
\put(0,#1){\vector(0,-1){#1}}%
\end{picture}}}\end{picture}}%

\newcommand{\SDOTAR}[1]%
{\truex{100}\truey{300}%
\NUMBEROFDOTS=#1%
\divide\NUMBEROFDOTS by \value{y}%
\begin{picture}(0,0)%
\put(0,0){\makebox(0,0)%
{\begin{picture}(0,#1)%
\multiput(0,#1)(0,-\value{y}){\NUMBEROFDOTS}%
{\circle*{\value{x}}}%
\put(0,0){\vector(0,-1){0}}%
\end{picture}}}\end{picture}}%

\newcommand{\SMONO}[1]%
{\begin{picture}(0,0)%
\put(0,0){\makebox(0,0)%
{\begin{picture}(0,#1)%
\put(0,#1){\vector(0,-1){#1}}%
\truex{300}\truey{600}%
\put(0,#1){\begin{picture}(0,0)%
\put(-\value{x},-\value{x}){\line(1,0){\value{y}}}\end{picture}}%
\end{picture}}}\end{picture}}%

\newcommand{\SEPI}[1]%
{\begin{picture}(0,0)%
\put(0,0){\makebox(0,0)%
{\begin{picture}(0,#1)%
\put(0,#1){\vector(0,-1){#1}}%
\truex{300}\truey{600}\truez{800}%
\put(-\value{x},\value{z}){\line(1,0){\value{y}}}%
\end{picture}}}\end{picture}}%

\newcommand{\SBIMO}[1]%
{\begin{picture}(0,0)%
\put(0,0){\makebox(0,0)%
{\begin{picture}(0,#1)%
\put(0,#1){\vector(0,-1){#1}}%
\truex{300}\truey{600}\truez{800}%
\put(0,#1){\begin{picture}(0,0)%
\put(-\value{x},-\value{x}){\line(1,0){\value{y}}}\end{picture}}%
\put(-\value{x},\value{z}){\line(1,0){\value{y}}}%
\end{picture}}}\end{picture}}%

\newcommand{\SBIAR}[1]%
{\begin{picture}(0,0)%
\truex{200}%
\put(0,0){\makebox(0,0)%
{\begin{picture}(0,#1)\put(-\value{x},#1){\vector(0,-1){#1}}%
\put(\value{x},#1){\vector(0,-1){#1}}%
\end{picture}}}\end{picture}}%

\newcommand{\SEQL}[1]%
{\begin{picture}(0,0)%
\truex{100}%
\put(0,0){\makebox(0,0)%
{\begin{picture}(0,#1)\put(-\value{x},#1){\line(0,-1){#1}}%
\put(\value{x},#1){\line(0,-1){#1}}%
\end{picture}}}\end{picture}}%

\newcommand{\Sarv}[2]{\Vcase{\SAR}{#1}{#200}}%

\newcommand{\Sar}[1]{\Sarv{#1}{50}}%

\newcommand{\saRv}[2]{\vcasE{\SAR}{#1}{#200}}%

\newcommand{\saR}[1]{\saRv{#1}{50}}%

\newcommand{\Sisov}[2]%
{\Vbicase{\SAR}{#1\hspace{-2pt}}{\hspace{-2pt}\cong}{#200}}%

\newcommand{\Sbiarv}[3]{\Vbicase{\SBIAR}{#1}{#2}{#300}}%

\newcommand{\Sbiar}[2]{\Sbiarv{#1}{#2}{50}}%

\newcommand{\NAR}[1]%
{\begin{picture}(0,0)%
\put(0,0){\makebox(0,0)%
{\begin{picture}(0,#1)\put(0,0){\vector(0,1){#1}}%
\end{picture}}}\end{picture}}%

\newcommand{\NDOTAR}[1]%
{\truex{100}\truey{300}%
\NUMBEROFDOTS=#1%
\divide\NUMBEROFDOTS by \value{y}%
\begin{picture}(0,0)%
\put(0,0){\makebox(0,0)%
{\begin{picture}(0,#1)%
\multiput(0,0)(0,\value{y}){\NUMBEROFDOTS}%
{\circle*{\value{x}}}%
\put(0,#1){\vector(0,1){0}}%
\end{picture}}}\end{picture}}%

\newcommand{\NMONO}[1]%
{\begin{picture}(0,0)%
\put(0,0){\makebox(0,0)%
{\begin{picture}(0,#1)%
\put(0,0){\vector(0,1){#1}}%
\truex{300}\truey{600}%
\put(-\value{x},\value{x}){\line(1,0){\value{y}}}%
\end{picture}}}%
\end{picture}}%

\newcommand{\NEPI}[1]%
{\begin{picture}(0,0)%
\put(0,0){\makebox(0,0)%
{\begin{picture}(0,#1)%
\put(0,0){\vector(0,1){#1}}%
\truex{300}\truey{600}\truez{800}%
\put(0,#1){\begin{picture}(0,0)%
\put(-\value{x},-\value{z}){\line(1,0){\value{y}}}\end{picture}}%
\end{picture}}}\end{picture}}%

\newcommand{\NBIMO}[1]%
{\begin{picture}(0,0)%
\put(0,0){\makebox(0,0)%
{\begin{picture}(0,#1)%
\put(0,0){\vector(0,1){#1}}%
\truex{300}\truey{600}\truez{800}%
\put(-\value{x},\value{x}){\line(1,0){\value{y}}}%
\put(0,#1){\begin{picture}(0,0)%
\put(-\value{x},-\value{z}){\line(1,0){\value{y}}}\end{picture}}%
\end{picture}}}\end{picture}}%

\newcommand{\NBIAR}[1]%
{\begin{picture}(0,0)%
\truex{200}%
\put(0,0){\makebox(0,0)%
{\begin{picture}(0,#1)\put(-\value{x},0){\vector(0,1){#1}}%
\put(\value{x},0){\vector(0,1){#1}}%
\end{picture}}}\end{picture}}%

\newcommand{\Narv}[2]{\Vcase{\NAR}{#1}{#200}}%

\newcommand{\Nar}[1]{\Narv{#1}{50}}%

\newcommand{\Nisov}[2]%
{\Vbicase{\NAR}{#1\hspace{-2pt}}{\hspace{-2pt}\cong}{#200}}%

\newcommand{\Nbiarv}[3]{\Vbicase{\NBIAR}{#1}{#2}{#300}}%

\newcommand{\Nbiar}[2]{\Nbiarv{#1}{#2}{50}}%

\newcommand{\NEDOTAR}%
{\truex{100}\truey{212}%
\NUMBEROFDOTS=5800%
\divide\NUMBEROFDOTS by \value{y}%
\begin{picture}(0,0)%
\multiput(-2900,-2900)(\value{y},\value{y}){\NUMBEROFDOTS}%
{\circle*{\value{x}}}%
\put(2900,2900){\vector(1,1){0}}%
\end{picture}}%

\newcommand{\SWDOTAR}%
{\truex{100}\truey{212}%
\NUMBEROFDOTS=5800%
\divide\NUMBEROFDOTS by \value{y}%
\begin{picture}(0,0)%
\multiput(2900,2900)(-\value{y},-\value{y}){\NUMBEROFDOTS}%
{\circle*{\value{x}}}%
\put(-2900,-2900){\vector(-1,-1){0}}%
\end{picture}}%

\newcommand{\sdcase}[3]{\begin{picture}(0,0)%
\put(0,-150){#1}%
\truex{100}\truez{600}%
\put(\value{x},\value{x}){\makebox(0,\value{z})[l]{${#2}$}}%
\truex{300}\truey{800}%
\put(-\value{x},-\value{y}){\makebox(0,\value{z})[r]{${#3}$}}%
\end{picture}}%

\newcommand{\SEDOTAR}%
{\truex{100}\truey{212}%
\NUMBEROFDOTS=5800%
\divide\NUMBEROFDOTS by \value{y}%
\begin{picture}(0,0)%
\multiput(-2900,2900)(\value{y},-\value{y}){\NUMBEROFDOTS}%
{\circle*{\value{x}}}%
\put(2900,-2900){\vector(1,-1){0}}%
\end{picture}}%

\newcommand{\Sedotar}[1]{\sdcase{\SEDOTAR}{#1}{}}%

\newcommand{\NWDOTAR}%
{\truex{100}\truey{212}%
\NUMBEROFDOTS=5800%
\divide\NUMBEROFDOTS by \value{y}%
\begin{picture}(0,0)%
\multiput(2900,-2900)(-\value{y},\value{y}){\NUMBEROFDOTS}%
{\circle*{\value{x}}}%
\put(-2900,2900){\vector(-1,1){0}}%
\end{picture}}%

\newcommand{\Nwdotar}[1]{\sdcase{\NWDOTAR}{#1}{}}%

\newcommand{\ENEAR}[2]%
{\makebox[0pt]{\begin{picture}(0,0)%
\put(0,-150){\makebox(0,0){\begin{picture}(0,0)%
\put(-6600,-3300){\vector(2,1){13200}}%
\truex{200}\truey{800}\truez{600}%
\put(-\value{x},\value{x}){\makebox(0,\value{z})[r]{${#1}$}}%
\put(\value{x},-\value{y}){\makebox(0,\value{z})[l]{${#2}$}}%
\end{picture}}}\end{picture}}}%

\newcommand{\ESEAR}[2]%
{\makebox[0pt]{\begin{picture}(0,0)%
\put(0,-150){\makebox(0,0){\begin{picture}(0,0)%
\put(-6600,3300){\vector(2,-1){13200}}%
\truex{200}\truey{800}\truez{600}%
\put(\value{x},\value{x}){\makebox(0,\value{z})[l]{${#1}$}}%
\put(-\value{x},-\value{y}){\makebox(0,\value{z})[r]{${#2}$}}%
\end{picture}}}\end{picture}}}%

\newcommand{\Esear}[1]{\ESEAR{#1}{}}%

\newcommand{\eseaR}[1]{\ESEAR{}{#1}}%

\newcommand{\WNWAR}[2]%
{\makebox[0pt]{\begin{picture}(0,0)%
\put(0,-150){\makebox(0,0){\begin{picture}(0,0)%
\put(6600,-3300){\vector(-2,1){13200}}%
\truex{200}\truey{800}\truez{600}%
\put(\value{x},\value{x}){\makebox(0,\value{z})[l]{${#1}$}}%
\put(-\value{x},-\value{y}){\makebox(0,\value{z})[r]{${#2}$}}%
\end{picture}}}\end{picture}}}%

\newcommand{\wnwaR}[1]{\WNWAR{}{#1}}%

\newcommand{\WSWAR}[2]%
{\makebox[0pt]{\begin{picture}(0,0)%
\put(0,-150){\makebox(0,0){\begin{picture}(0,0)%
\put(6600,3300){\vector(-2,-1){13200}}%
\truex{200}\truey{800}\truez{600}%
\put(-\value{x},\value{x}){\makebox(0,\value{z})[r]{${#1}$}}%
\put(\value{x},-\value{y}){\makebox(0,\value{z})[l]{${#2}$}}%
\end{picture}}}\end{picture}}}%

\newcommand{\NNEAR}[2]%
{\raisebox{-1pt}[0pt][0pt]{\begin{picture}(0,0)%
\put(0,0){\makebox(0,0){\begin{picture}(0,0)%
\put(-3300,-6600){\vector(1,2){6600}}%
\truex{100}\truez{600}%
\put(-\value{x},\value{x}){\makebox(0,\value{z})[r]{${#1}$}}%
\put(\value{x},-\value{z}){\makebox(0,\value{z})[l]{${#2}$}}%
\end{picture}}}\end{picture}}}%

\newcommand{\SSWAR}[2]%
{\raisebox{-1pt}[0pt][0pt]{\begin{picture}(0,0)%
\put(0,0){\makebox(0,0){\begin{picture}(0,0)%
\put(3300,6600){\vector(-1,-2){6600}}%
\truex{100}\truez{600}%
\put(-\value{x},\value{x}){\makebox(0,\value{z})[r]{${#1}$}}%
\put(\value{x},-\value{z}){\makebox(0,\value{z})[l]{${#2}$}}%
\end{picture}}}\end{picture}}}%

\newcommand{\SSEAR}[2]%
{\raisebox{-1pt}[0pt][0pt]{\begin{picture}(0,0)%
\put(0,0){\makebox(0,0){\begin{picture}(0,0)%
\put(-3300,6600){\vector(1,-2){6600}}%
\truex{200}\truez{600}%
\put(\value{x},\value{x}){\makebox(0,\value{z})[l]{${#1}$}}%
\put(-\value{x},-\value{z}){\makebox(0,\value{z})[r]{${#2}$}}%
\end{picture}}}\end{picture}}}%

\newcommand{\Ssear}[1]{\SSEAR{#1}{}}%

\newcommand{\sseaR}[1]{\SSEAR{}{#1}}%

\newcommand{\NNWAR}[2]%
{\raisebox{-1pt}[0pt][0pt]{\begin{picture}(0,0)%
\put(0,0){\makebox(0,0){\begin{picture}(0,0)%
\put(3300,-6600){\vector(-1,2){6600}}%
\truex{200}\truez{600}%
\put(\value{x},\value{x}){\makebox(0,\value{z})[l]{${#1}$}}%
\put(-\value{x},-\value{z}){\makebox(0,\value{z})[r]{${#2}$}}%
\end{picture}}}\end{picture}}}%

\newcommand{\Nnwar}[1]{\NNWAR{#1}{}}%

\newcommand{\Necurve}[2]%
{\begin{picture}(0,0)%
\truex{1300}\truey{2000}\truez{200}%
\put(0,\value{x}){\oval(#200,\value{y})[t]}%
\put(0,\value{x}){\makebox(0,0){\begin{picture}(#200,0)%
\put(#200,0){\vector(0,-1){\value{z}}}%
\put(0,0){\line(0,-1){\value{z}}}\end{picture}}}%
\truex{2500}%
\put(0,\value{x}){\makebox(0,0)[b]{${#1}$}}%
\end{picture}}%

\newcommand{\Nwcurve}[2]%
{\begin{picture}(0,0)%
\truex{1300}\truey{2000}\truez{200}%
\put(0,\value{x}){\oval(#200,\value{y})[t]}%
\put(0,\value{x}){\makebox(0,0){\begin{picture}(#200,0)%
\put(#200,0){\line(0,-1){\value{z}}}%
\put(0,0){\vector(0,-1){\value{z}}}\end{picture}}}%
\truex{2500}%
\put(0,\value{x}){\makebox(0,0)[b]{${#1}$}}%
\end{picture}}%

\newcommand{\Securve}[2]%
{\begin{picture}(0,0)%
\truex{1300}\truey{2000}\truez{200}%
\put(0,-\value{x}){\oval(#200,\value{y})[b]}%
\put(0,-\value{x}){\makebox(0,0){\begin{picture}(#200,0)%
\put(#200,0){\vector(0,1){\value{z}}}%
\put(0,0){\line(0,1){\value{z}}}\end{picture}}}%
\truex{2500}%
\put(0,-\value{x}){\makebox(0,0)[t]{${#1}$}}%
\end{picture}}%

\newcommand{\Swcurve}[2]%
{\begin{picture}(0,0)%
\truex{1300}\truey{2000}\truez{200}%
\put(0,-\value{x}){\oval(#200,\value{y})[b]}%
\put(0,-\value{x}){\makebox(0,0){\begin{picture}(#200,0)%
\put(#200,0){\line(0,1){\value{z}}}%
\put(0,0){\vector(0,1){\value{z}}}\end{picture}}}%
\truex{2500}%
\put(0,-\value{x}){\makebox(0,0)[t]{${#1}$}}%
\end{picture}}%

\newcommand{\Escurve}[2]%
{\begin{picture}(0,0)%
\truex{1400}\truey{2000}\truez{200}%
\put(\value{x},0){\oval(\value{y},#200)[r]}%
\put(\value{x},0){\makebox(0,0){\begin{picture}(0,#200)%
\put(0,0){\vector(-1,0){\value{z}}}%
\put(0,#200){\line(-1,0){\value{z}}}\end{picture}}}%
\truex{2500}%
\put(\value{x},0){\makebox(0,0)[l]{${#1}$}}%
\end{picture}}%

\newcommand{\Encurve}[2]%
{\begin{picture}(0,0)%
\truex{1400}\truey{2000}\truez{200}%
\put(\value{x},0){\oval(\value{y},#200)[r]}%
\put(\value{x},0){\makebox(0,0){\begin{picture}(0,#200)%
\put(0,0){\line(-1,0){\value{z}}}%
\put(0,#200){\vector(-1,0){\value{z}}}\end{picture}}}%
\truex{2500}%
\put(\value{x},0){\makebox(0,0)[l]{${#1}$}}%
\end{picture}}%

\newcommand{\Wscurve}[2]%
{\begin{picture}(0,0)%
\truex{1300}\truey{2000}\truez{200}%
\put(-\value{x},0){\oval(\value{y},#200)[l]}%
\put(-\value{x},0){\makebox(0,0){\begin{picture}(0,#200)%
\put(0,0){\vector(1,0){\value{z}}}%
\put(0,#200){\line(1,0){\value{z}}}\end{picture}}}%
\truex{2400}%
\put(-\value{x},0){\makebox(0,0)[r]{${#1}$}}%
\end{picture}}%

\newcommand{\Wncurve}[2]%
{\begin{picture}(0,0)%
\truex{1300}\truey{2000}\truez{200}%
\put(-\value{x},0){\oval(\value{y},#200)[l]}%
\put(-\value{x},0){\makebox(0,0){\begin{picture}(0,#200)%
\put(0,0){\line(1,0){\value{z}}}%
\put(0,#200){\vector(1,0){\value{z}}}\end{picture}}}%
\truex{2400}%
\put(-\value{x},0){\makebox(0,0)[r]{${#1}$}}%
\end{picture}}%

\newcount\SCALE%

\newcount\NUMBER%

\newcount\LINE%

\newcount\COLUMN%

\newcount\WIDTH%

\newcount\SOURCE%

\newcount\ARROW%

\newcount\TARGET%

\newcount\ARROWLENGTH%

\newcount\NUMBEROFDOTS%

\newcounter{x}%

\newcounter{y}%

\newcounter{z}%

\newcounter{horizontal}%

\newcounter{vertical}%

\newskip\itemlength%

\newskip\firstitem%

\newskip\seconditem%

\newcommand{\printarrow}{}%

\newcommand{\truex}[1]{%
\NUMBER=#1%
\multiply\NUMBER by 100%
\divide\NUMBER by \SCALE%
\setcounter{x}{\NUMBER}}%

\newcommand{\truey}[1]{%
\NUMBER=#1%
\multiply\NUMBER by 100%
\divide\NUMBER by \SCALE%
\setcounter{y}{\NUMBER}}%

\newcommand{\truez}[1]{%
\NUMBER=#1%
\multiply\NUMBER by 100%
\divide\NUMBER by \SCALE%
\setcounter{z}{\NUMBER}}%

\newcommand{\changecounters}[1]{%
\SOURCE=\ARROW%
\ARROW=\TARGET%
\settowidth{\itemlength}{#1}%
\ifdim \itemlength > 2800\unitlength%
\addtolength{\itemlength}{-2800\unitlength}%
\TARGET=\itemlength%
\divide\TARGET by 1310%
\multiply\TARGET by 100%
\divide\TARGET by \SCALE%
\else%
\TARGET=0%
\fi%
\ARROWLENGTH=5000%
\advance\ARROWLENGTH by -\SOURCE%
\advance\ARROWLENGTH by -\TARGET%
\advance\SOURCE by -\TARGET}%

\newcommand{\initialize}[1]{%
\LINE=0%
\COLUMN=0%
\WIDTH=0%
\ARROW=0%
\TARGET=0%
\changecounters{#1}%
\renewcommand{\printarrow}{#1}%
\begin{center}%
\vspace{10pt}%
\begin{picture}(0,0)}%

\newcommand{\DIAGV}[2]{%
\SCALE=#1%
\setlength{\unitlength}{655sp}%
\multiply\unitlength by \SCALE%
\divide\unitlength by 100%
\initialize{\mbox{$#2$}}}%

\newcommand{\n}[1]{%
\changecounters{\mbox{$#1$}}%
\put(\COLUMN,\LINE){\makebox(0,0){\printarrow}}%
\thinlines%
\renewcommand{\printarrow}{\mbox{$#1$}}%
\advance\COLUMN by 4000}%

\newcommand{\nn}[1]{%
\put(\COLUMN,\LINE){\makebox(0,0){\printarrow}}%
\thinlines%
\ifnum \WIDTH < \COLUMN%
\WIDTH=\COLUMN%
\else%
\fi%
\advance\LINE by -4000%
\COLUMN=0%
\ARROW=0%
\TARGET=0%
\changecounters{\mbox{$#1$}}%
\renewcommand{\printarrow}{\mbox{$#1$}}}%

\newcommand{\conclude}{%
\put(\COLUMN,\LINE){\makebox(0,0){\printarrow}}%
\thinlines%
\ifnum \WIDTH < \COLUMN%
\WIDTH=\COLUMN%
\else%
\fi%
\setcounter{horizontal}{\WIDTH}%
\setcounter{vertical}{-\LINE}%
\end{picture}}%

\newcommand{\diag}{%
\conclude%
\raisebox{0pt}[0pt][\value{vertical}\unitlength]{}%
\hspace*{\value{horizontal}\unitlength}%
\vspace{10pt}%
\end{center}%
\setlength{\unitlength}{1pt}}%

\newcommand{\diagv}[3]{%
\conclude%
\NUMBER=#1%
\rule{0pt}{\NUMBER pt}%
\hspace*{-#2pt}%
\raisebox{0pt}[0pt][\value{vertical}\unitlength]{}%
\hspace*{\value{horizontal}\unitlength}%6
\NUMBER=#3%
\advance\NUMBER by 10%
\vspace*{\NUMBER pt}%
\end{center}%
\setlength{\unitlength}{1pt}}%

\newcommand{\N}[1]%
{\raisebox{0pt}[7pt][0pt]{$#1$}}%

\newcommand{\crosslength}[2]{%
\settowidth{\firstitem}{#1}%
\settowidth{\seconditem}{#2}%
\ifdim\firstitem < \seconditem%
\itemlength=\seconditem%
\else%
\itemlength=\firstitem%
\fi%
\divide\itemlength by 2%
\hspace{\itemlength}}%

\newcommand{\B}{\thicklines}

\def\nat{\raisebox{-2pt}%
{\begin{picture}(34,10)%
\put(2,10){\vector(1,0){30}}%
\put(2,0){\vector(1,0){30}}%
\put(13,2){$\Downarrow$}%
\end{picture}}}%